\documentclass[%
 reprint,
 amsmath,amssymb,
 aps,
]{revtex4-2}

\usepackage{graphicx}
\usepackage{dcolumn}
\usepackage{bm}
\usepackage{textcomp} 

\usepackage[version=3]{mhchem}

\begin{document}

\preprint{APS/123-QED}

\title{Multiscale Modeling of Metal/Oxide/Metal Conductive Bridging Random Access Memory Cells: from \textit{Ab Initio} to Finite Element Calculations}

\author{Jan Aeschlimann}
\email{aejan@iis.ee.ethz.ch}
\author{Fabian Ducry}%
\author{Christoph Weilenmann}
\author{Alexandros Emboras}
\author{Mathieu Luisier}
\affiliation{%
 Integrated Systems Laboratory, ETH Z\"urich\\
 Gloriastrasse 35, 8092 Z\"urich, Switzerland
}%

\author{Juerg Leuthold}
\affiliation{
 Institute of Electromagnetic Fields, ETH Z\"urich\\
 Gloriastrasse 35, 8092 Z\"urich, Switzerland
}%


\date{\today}

\begin{abstract}
We present a multiscale simulation framework to compute the current vs.\ voltage (\textit{I-V}) characteristics of metal/oxide/metal structures building the core of conductive bridging random access memory (CBRAM) cells and to shed light on their resistance switching properties. The approach relies on a finite element model whose input material parameters are extracted either from \textit{ab initio} or from machine-learned empirical calculations. The applied techniques range from molecular dynamics and nudged elastic band to electronic and thermal quantum transport. Such an approach drastically reduces the number of fitting parameters needed and makes the resulting modeling environment more accurate than traditional ones. The developed computational framework is then applied to the investigation of an Ag/a-\ce{SiO2}/Pt CBRAM, reproducing experimental data very well. Moreover, the relevance of Joule heating is assessed by considering various cell geometries. It is found that self-heating manifests itself in devices with thin conductive filaments with few-nanometer diameters and at current concentrations in the 10s-microampere range. With the proposed methodology it is now possible to explore the potential of not-yet fabricated memory cells and to reliably optimize their design. 
\end{abstract}

\maketitle

\section{\label{sec:introduction}Introduction}

Driven by Moore's scaling law and the introduction of innovative technology boosters, the speed of microprocessors has increased much faster than that of storage units, giving rise to the well-known "memory wall" issue in computing \cite{Wulf1995}. Well-established memory technologies such as Flash and dynamic random access memory (DRAM) cells are reaching their limit both in terms of dimensions scaling and integration density so that the three-dimensional (3-D) stacking of multiple memory layers has already started \cite{Zidan2018,Wong2015}. Besides the speed and scaling issues, the energy efficiency of individual memory cells is becoming an important aspect as well. Especially, the non-volatile storage of information might pave the way for novel applications such as edge \cite{Chen2019}, neuromorphic \cite{Hong2018,Sokolov2021}, or in-memory \cite{Ielmini2018,Pedretti2021} computing where the energy consumption must be minimal. Therefore, several memory technologies have recently emerged \cite{Waser2009,Gupta2020,Zahoor2020}. 

\begin{figure*}
\includegraphics[width=17.6cm]{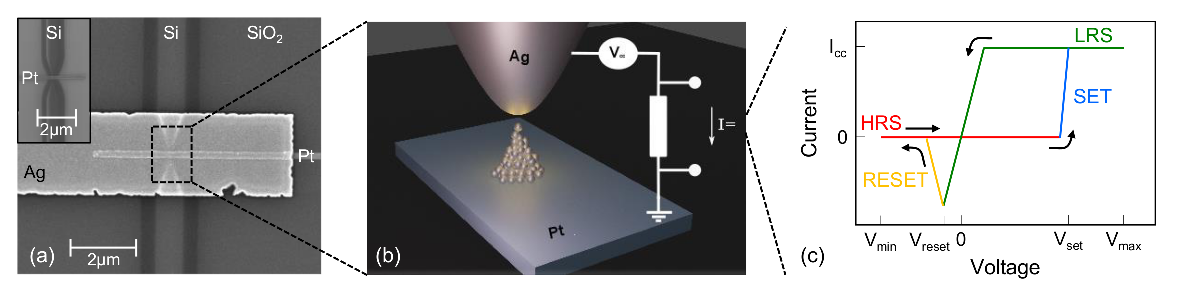}
\caption{\label{fig:fig_experiment} (a) Scanning electron microscope view of an Ag/a-SiO$_2$/Pt CBRAM cell that was fabricated on a silicon-on-insulator (SOI) wafer \cite{Emboras2018}. The bottom and top contacts are Pt and Ag, respectively. They surround a 20~nm thick layer of amorphous SiO$_2$ (a-SiO$_2$). The inset shows the bottom Pt electrode, including a buried Si waveguide to confine the active switching region. (b) Illustration of the active region of a typical CBRAM structure with Pt and Ag contacts. A DC voltage is applied to the Ag contact; the Pt one remains grounded. This bias triggers the growth of a nanoscale Ag filament starting from the Pt side, eventually bridging the two contacts through the SiO$_2$ network (not shown in this sub-plot). (c) Typical "current vs. voltage" (\emph{I-V}) characteristics of a non-volatile CBRAM as in (a) and (b). Starting from the high-resistance state (HRS), the cell switches into its low-resistance state (LRS) at $V_\text{set}$. A compliance current $I_{cc}$ is applied to avoid current-driven damages. The RESET back to the HRS occurs at $V_\text{reset}$, which must be negative for non-volatile storage. }
\end{figure*}

One promising candidate that could address all aforementioned challenges is the conductive bridging random access memory (CBRAM) technology, also known as electrochemical metallization (ECM) cells. The most primitive form of CBRAM consists of a capacitor-like metal/solid electrolyte/metal structure that switches between a high- (HRS) and a low-resistance state (LRS) through the reversible growth and dissolution of a conductive metallic filament \cite{Waser2009}. This process occurs within the solid electrolyte, typically an insulating layer, upon application of an external voltage, as can be seen in Fig.~\ref{fig:fig_experiment}(a)-(b). CBRAMs are characterized by an extremely low power consumption (write energy down to $\sim 0.1$~pJ/bit \cite{Lanza2022}), a large high-to-low resistance ratio ($>10^5$ \cite{Belmonte2021}), and excellent scalability with cross sections smaller than $50 \times 50$~nm$^2$ \cite{Cheng2019}. Furthermore, CBRAM cells allow for non-volatile storage so that programmed states are retained for long times ($>10$ years \cite{Lanza2022}), even when the supply voltage is turned off. Their "current vs.\ voltage" (\emph{I-V}) characteristics exhibit a hysteretic behavior and comprises a SET and RESET phase between its two distinct resistance states, LRS and HRS, as depicted in Fig.~\ref{fig:fig_experiment}(c). Lastly, the fabrication process and the materials composing these memories can be made CMOS compatible, which enables CBRAMs to be readily integrated into today's microprocessors \cite{Ielmini2016}. 

A wide range of electrode and insulator materials have been combined with each other to enhance the CBRAM performance \cite{Wang2018}. To drive this technology towards its limits, the device sizes have been scaled down such that only few atoms are involved in their HRS-to-LRS switching process \cite{Cheng2019}. This downsizing towards the atomic scale, however, leads to microscopic differences in the atomic arrangements of different devices and cycles. The fact that resistance switching may only rely on few atoms might negatively impact the reliability and retention times of CBRAM cells \cite{Zahoor2020}. 

To support the on-going experimental activity on CBRAM, physics-based simulation of such devices has gained in importance. Required are modeling tools that can shed light on the operational principle of CBRAMs while accounting for the atomic scale variation that occurs from cell to cell or even from cycle to cycle. \emph{Ab initio} quantum mechanical modeling methods such as density functional theory (DFT) \cite{Kohn1965} can describe the electronic structure of any given atomic configuration. Thanks to their high accuracy and versatility, they are frequently used to derive the material properties of CBRAM structures \cite{Pandey2015,Sankaran2012,Xiao2016}. With \emph{ab initio} molecular dynamics (AIMD), it is also possible to study chemical processes at the atomic level, e.g., the formation of metallic clusters during the SET process \cite{Akola2022}. However, the high computational costs limit such studies to time ranges of a few hundred picoseconds and structures comprising of ca.\ 1000 atoms. Since even the fastest reported CBRAM cells exhibit switching times in the order of a few nanoseconds \cite{Geresdi2014}, full SET/RESET processes of entire CBRAM cells cannot be studied with \emph{ab initio} methods. On the other hand, a complete CBRAM switching cycle at the atomic level was demonstrated using classical MD based on semi-empirical force-fields (FF) approaches \cite{Onofrio2015,Wang2019p1,Wang2019nat,Pandey2015}. In such simulations, a parameter set describes the different types of atoms and their interactions. The parameters are fitted to reproduce reference data typically from quantum mechanical calculations and are tailored to a specific material system. Although recent approaches using machine-learning algorithms \cite{Podryabinkin2017,Zuo2020} have facilitated the fitting of the required parameters, the generation and validation of force-field parameter sets remains challenging and computationally expensive. Furthermore, the SET and RESET processes can typically only be triggered by applying voltage pulses of constant biases with \emph{ab initio} and FF approaches. 

To simulate full hysteretic CBRAM switching cycles following a voltage sweep similar to the one in Fig.~\ref{fig:fig_experiment}(c), the kinetic Monte Carlo (KMC) method can be applied \cite{Menzel2015,Larcher2017,Dirkmann2017}. The KMC simulation domain is typically discretized into a grid with fixed atom sites. The movement of atoms between the sites is governed by rate equations obeying Arrhenius-type relations. It is generally challenging to determine realistic material parameters for these rate equations, often leading to a substantial amount of fitting parameters that limits the accuracy of such models. Furthermore, it is difficult to account for temperature effects and the current is often computed by using rudimentary network models that analytically look for a connecting path between the two electrodes \cite{Onofrio2015}. 

As an alternative to KMC, continuum models \cite{Menzel2017,Wang2019p1,Wang2019nat,Lin2012,Lubben2017,Menzel2013,Russo2009,Wang2019p2,Larentis2012,Dorion2013,Ielmini2011} have been proposed to investigate the switching process of CBRAM cells. They mostly rely on the classical physics and consist of differential equations that describe the underlying physical phenomena, e.g., ion transport, electrostatics, or heat generation. Their underlying solvers take advantage of a broad range of modules that describe the physical phenomena and that can be selected and combined with each other. The resulting set of equations must be typically solved self-consistently. In contrast to the aforementioned methods, continuum models do not capture the atomic granularity of the metallic filaments nor the amorphous nature of the insulating layers. The loss in accuracy is thus compensated by a high computational efficiency. Finally, continuum models provide a high versatility in terms of time scales (from picoseconds to hours), sample size (from nanometer to meter), and sample geometry. Continuum models can produce full \emph{I-V} cycles so that the described data can be directly compared to experiments. 

Examples of continuum models are those relying on the finite element method (FEM). In case of CBRAM cells, they generally comprise an active top electrode, a solid electrolyte acting as switching layer, and an inert bottom electrode. All regions are discretized with finite elements. The description of the growth and dissolution of the metal filament within the insulator represents the key kernel of any CBRAM switching model. To do so, the most relevant electrochemical and physical processes must be taken into account, namely redox reactions (oxidation/reduction) at the metal/solid electrolyte interfaces and metal ion migration through the solid electrolyte. To circumvent the physical description of the nucleation phase during the filament formation, a truncated metal filament attached to the inert electrode is often inserted into the switching layer as a seed. The model of Ref.~\cite{Menzel2017} is an excellent illustration of this approach. It describes both the SET and RESET operations of CBRAMs by using a deformed two-dimensional (2-D) geometry algorithm that iteratively moves the filament boundary. The full \emph{I-V} characteristics are then calculated with the help of analytical formulas. Besides FEM, the level-set method, which takes advantage of a fixed instead of moving mesh, is also frequently employed, as proposed in Ref.~\cite{Lin2012}. There, the metal/solid electrolyte interface is treated more explicitly by including a Helmholtz double layer. Additionally, this model can be implemented in 3-D by assuming rotational symmetry so that more precise filament shapes can be examined. However, the electrical current flowing through these structures cannot be directly extracted. To reduce the complexity of Refs.~\cite{Menzel2017} and \cite{Lin2012}, alternative models have been developed that can only simulate the forming/SET \cite{Lubben2017} or the RESET step \cite{Wang2019nat,Menzel2013,Russo2009}. Focusing only on the RESET operation is of particular interest as it is influenced by Joule heating: it has been shown that the filament dissolution can be thermally assisted \cite{Menzel2013,Menzel2017,Wang2019p2,Larentis2012,Russo2009}. The numerical model of Ref.~\cite{Larentis2012}, for example, accounts for Joule heating, featuring the temperature evolution of the conductive filament during the RESET process and its impact on the \emph{I-V} characteristics. Also, this model provides a physical insight into the microscopic morphology of the filament in the low- and high-resistance states. 

All continuum models generally suffer from the same limitation: the parameters needed to describe the involved electrochemical, ionic, electronic, and thermal processes are either derived from experiments or serve as fitting quantities. This considerably limits their predictive power. Due to the large number of required material parameters, typically more than ten depending on the complexity of the model, it is very challenging to derive all of them from experiments, thus preventing the exploration of entirely novel device concepts or material combinations. Importantly, the increasing structural complexity and the decreasing dimensions of the currently fabricated devices call for an investigation of the material properties at the atomic level. At this scale, the reported bulk values might no more be exact and should be replaced by system-specific parameters \cite{Martin2022}. 

In this work, we propose a multiscale simulation platform where the advantages of some of the aforementioned methods are combined: The versatility and computational efficiency of FEM models are leveraged to capture the geometrical features of CBRAMs, while the required input parameters are determined through \emph{ab initio} and force-field calculations. This avoids the usage of sometimes complex fitting procedures. Specifically, the essential parameters describing the physical processes occurring in CBRAM cells are computed at the atomic level by using either \emph{ab initio} methods or, for the thermal aspects, semi-empirical parameters trained on first-principles calculations. The result is an advanced, multiscale tool chain where parameters are extracted from DFT calculations using AIMD, nudged elastic band (NEB), and electronic and thermal quantum transport simulations. As an illustration, the modeling environment is applied to a silver/amorphous silicon dioxide/platinum (Ag/a-SiO$_2$/Pt) CBRAM cell, as reported in Ref.~\cite{Emboras2018}. We demonstrate that the experimental \emph{I-V} characteristics are well reproduced with almost no fitting parameters. Finally, we assess the role of Joule heating by varying the original CBRAM structure and highlight internal temperature behavior as a function of the filament geometry and the drive current magnitude. 

The paper is organized as follows: the modules of the FEM model are discussed in Sec.~\ref{sec:fem}, stressing out the necessity of accurate input parameters. The evaluation of the diffusion and tunneling current parameters are described in Sec.~\ref{sec:diffusion} and \ref{sec:tunneling}, respectively, while the determination of the reaction barriers at the metal/solid electrolyte interfaces is introduced in Sec.~\ref{sec:fluxbarrier}. The approach to extract the electrical and thermal conductivity of the metallic filament is presented in Sec.~\ref{sec:conductivities}. Section~\ref{sec:result_iv} summarizes the key features and outputs of the implemented model with a concrete example. The relevance of Joule heating is examined in Sec.~\ref{sec:result_joule}. Finally, conclusions are drawn in Sec.~\ref{sec:conclusion}.

\section{\label{sec:approach}Modeling Approach}

\subsection{Finite Element Model}\label{sec:fem}

Our multiscale simulation environment is made of different components. In this subsection, the features of the FEM part, which derives inspiration from Refs.~\cite{Menzel2017,Lin2012}, its underlying physics, and its implementation are discussed. The FEM model is capable of providing full SET and RESET cycles of CBRAM cells including their corresponding \emph{I-V} characteristics. The time evolution of the filament shape during these cycles gives insights into the filament growth and dissolution process. Furthermore, the model sheds light on the relevance of Joule heating in CBRAM devices by capturing electro-thermal effects. The discussion of the FEM part is followed by a description of the modules to extract the material parameters governing the diffusion, chemical, electronic, and thermal processes with an Ag/a-SiO$_2$/Pt CBRAM cell as an example.

\begin{figure}
\includegraphics[width=8.0cm]{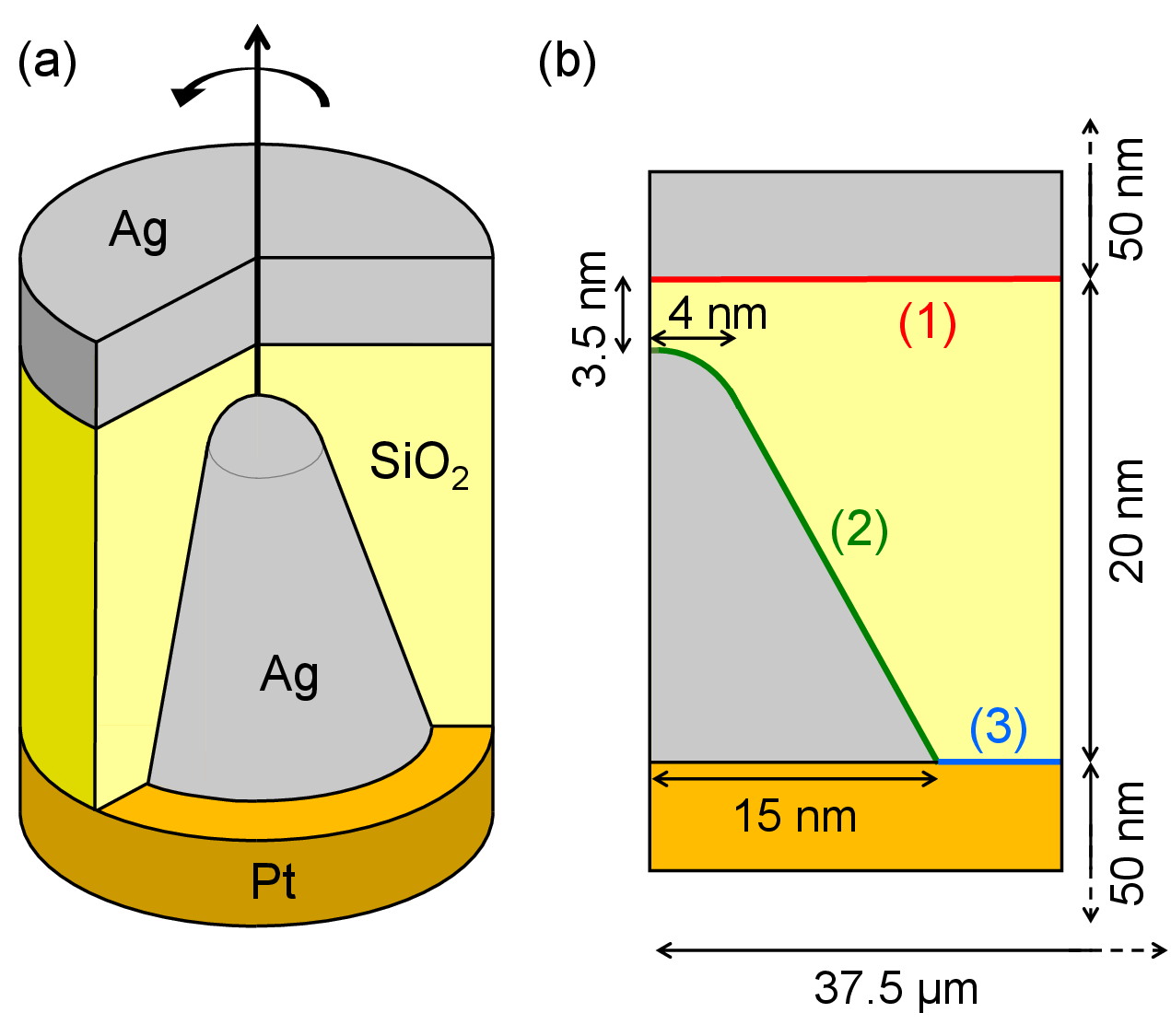}
\caption{\label{fig:fig_model} (a) Schematic view of the axially symmetric CBRAM geometry considered in the FEM model. In case of an Ag/a-SiO$_2$/Pt cell, the active top electrode is made of an Ag slab, the ion-conducting solid electrolyte layer consists of an a-SiO$_2$ switching layer, and a Pt slab acts as inert counter electrode. A moving cone-shaped Ag filament seed is initially placed within the switching layer. (b) Corresponding 2-D simulation domain under the assumption of rotational symmetry indicating the interfaces between the a-SiO$_2$ switching layer and (1) the Ag electrode, (2) the moving Ag filament, and (3) the Pt electrode. The dimensions of the structure are indicated by double arrows and are inspired by the experimental CBRAM cell from Ref.~\cite{Emboras2018}. }
\end{figure}

The CBRAM geometry as implemented in the FEM model is made of a metal/solid electrolyte/metal stack. The top and bottom electrodes are electrochemically active and inert, respectively, while the ion-conducting, but electrically insulating dielectric layer is placed in between. A schematic of the Ag/a-SiO$_2$/Pt structure considered in this work is shown in Fig.~\ref{fig:fig_model}. In addition, a truncated cone-shaped seed of the same material as the active electrode is attached to the inert electrode to enable the growth of a metallic nano-filament. Rotational symmetry along the filament axis is applied to reduce the simulation complexity from 3-D to quasi 2-D. 

The switching between the low- and high-resistance state of CBRAM cells is achieved by electrochemically growing and dissolving a conductive metal filament in the switching layer. When a positive voltage is applied to the anode of a cell in its HRS (interface (1) in Fig.~\ref{fig:fig_model}(b)), an oxidation reaction takes place. Metal ions are dissolved in the electrically insulating solid electrolyte layer. These cations migrate along the electric field towards the counter electrode (interface (2) and (3) in Fig.~\ref{fig:fig_model}(b)) where they are reduced and deposited as bulk metal. This growth process towards the anode continues as long as a voltage is applied and until the gap between the filament tip and the anode becomes short enough so that quantum mechanical tunneling produces a substantial electrical current. Eventually, the filament bridges (short-circuits) the two electrodes, giving rise to the device low-resistance state. To prevent damaging the cell the current is usually limited to a maximum compliance value, called $I_{cc}$. By applying a negative voltage, the process is reversed and the device switches back to its high-resistance state. 

\begin{figure*}
\includegraphics[width=17.6cm]{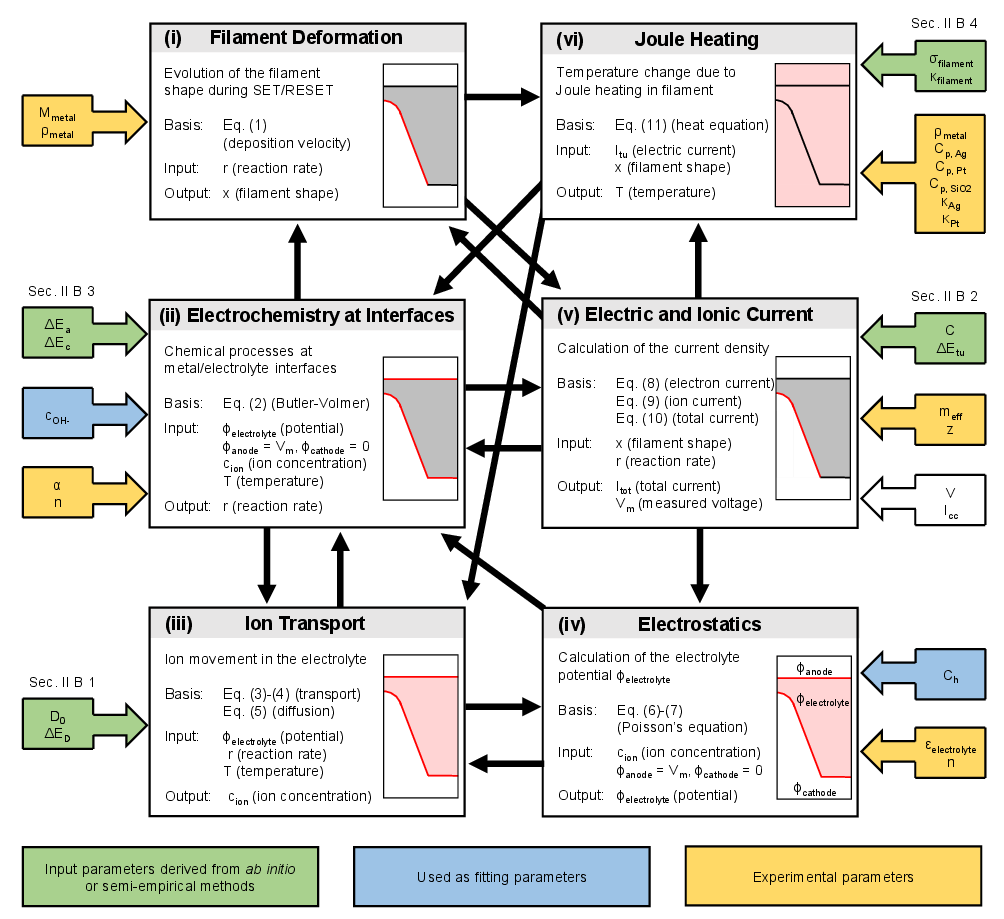}
\caption{\label{fig:workflow} Summary of the six modules implemented in the FEM model. Each box represents one module (anti-clockwise starting top left: (i) deformation, (ii) electrochemistry, (iii) transport, (iv) electrostatics, (v) currents, and (vi) Joule heating). It includes the underlying physical equation(s) in Sec.~\ref{sec:fem}, as well as the input and output parameters. All time-dependent  variables that are calculated by the model are listed in the boxes, while the material parameters are given on the side in colored arrows. The material parameters can either be derived from \emph{ab initio} and semi-empirical methods (green boxes with a reference to the corresponding section in the manuscript), used as fitting parameters (blue), or derived from experiments (yellow). The applied voltage ($V$) and the compliance current ($I_{cc}$) are taken from the experimental measurement of Sec.~\ref{sec:result_iv} (white). The dependence between the different modules is marked as black arrows that indicate the flow direction of the exchanged data. The insets in the boxes show a schematic of the modeled geometry following Fig.~\ref{fig:fig_model}(b). The regions to which the respective module is applied are colored red. 
}
\end{figure*}

To capture all physical processes that are relevant to the operation of CBRAM cells, the FEM model is split into six interdependent modules, as summarized in Fig.~\ref{fig:workflow}. Three processes are essential to describe the underlying physics of the SET and RESET operation of CBRAM cells: (i) the electrochemical, (ii) the ion transport, and (iii) the electrostatics module. The first one includes the oxidation and reduction reactions taking place at the metal/solid electrolyte interfaces. They are described by the Butler-Volmer equation, which provides a net reaction rate $r$. The second module, ion transport, simulates the migration of the metal ions through the solid electrolyte based on the drift-diffusion equation. The final module (electrostatics) calculates the potential of the solid electrolyte $\phi_\text{electrolyte}$ through Poisson's equation. These three modules are self-consistently coupled via $\phi_\text{electrolyte}$ and $r$. The size and shape of the filament at a specific time are determined by a fourth, deformation module. One of the key features of the FEM model resides in its capability to generate the full \emph{I-V} characteristics of the studied CBRAM devices. Hence, a fifth module to calculate both the electric and ionic currents must be implemented. Finally, Joule heating is captured by a sixth module. 

All six modules of the FEM model require material parameters as input. These material parameters can be divided into three categories. The first one is made of parameters derived from experimental studies of typically bulk structures, e.g., the density or the specific heat capacity. The second category comprises the parameters that are computed with \emph{ab initio} or semi-empirical methods. They include device-specific quantities such as the electrical or thermal conductivity of a nanoscale metal filament embedded in a solid electrolyte, which are challenging to measure experimentally. Instead, these parameters can be determined by the four modules described in Secs.~\ref{sec:diffusion}-\ref{sec:conductivities}. They govern the diffusion, electronic, chemical, and thermal processes of CBRAM cells. Lastly, the remaining parameters are set through fitting. They are typically chosen such that the model output fits experimental reference data. Their values must nevertheless be physically meaningful. 

The FEM model is implemented in COMSOL Multiphysics version 5.5 \cite{Comsol} assuming a rotationally symmetric, quasi 2-D Ag/a-SiO$_2$/Pt CBRAM structure, as displayed in Fig.~\ref{fig:fig_model}. Time-dependent simulations were performed to model the switching cycles with the same measurement settings as in the reference experiment that is detailed in Sec.~\ref{sec:result_iv}. The implicit backward differentiation formula (BDF) method \cite{Curtis1952} was used for time stepping. The entire structure is meshed into triangular elements with different granularity for computational efficiency. The finest mesh is generated along the filament/solid electrolyte interface with element sizes from 43~pm to 344~pm, while the elements of the remaining solid electrolyte can measure up to 2~nm. 

To simulate the filament growth and dissolution a moving mesh algorithm with Laplace mesh smoothing as implemented in COMSOL Multiphysics is applied within the deformation module of the FEM model (inset (i) of Fig.~\ref{fig:workflow}). The filament/solid electrolyte interface depicted as interface (2) in Fig.~\ref{fig:fig_model}(b) moves according to the normal mesh velocity $v_\text{dep}$, which is defined as \cite{Menzel2017} 
\begin{equation} \label{eq:vdep}
    v_\text{dep} = -\frac{M_\text{metal}}{\rho_\text{metal}} r. 
\end{equation}
In Eq.~(\ref{eq:vdep}), $M_\text{metal}$ and $\rho_\text{metal}$ are the molar mass and the density of the filament metal (i.e.\ Ag in the structure shown in Fig.~\ref{fig:fig_model}), respectively, while $r$ denotes the net reaction rate and will be detailed below. At all other boundaries, i.e., interfaces (1) and (3) in Fig.~\ref{fig:fig_model}(b), the normal mesh velocity is set to zero. When the filament has grown such that the compliance current value is reached, $v_\text{dep}$ is set to zero until $r$ switches polarity. During the filament growth/dissolution process the mesh is constantly deformed, which lowers its quality. Therefore, remeshing steps are performed to ensure numerical stability when the minimum element quality as defined under the deformed geometry node of COMSOL Multiphysics becomes below 0.2. For the geometry in Fig.~\ref{fig:fig_model}, about five remeshing steps are needed per switching cycle. 

The net reaction rate $r$ describes the electrochemical reaction kinetics at the metal/solid electrolyte interfaces (1)-(3) in Fig.~\ref{fig:fig_model}(b), i.e.\ $\text{Ag} \rightleftharpoons \text{Ag}^+ + \text{e}^-$ with e$^-$ being an electron. At these interfaces, two opposing reactions take place at the same time. On the one hand, interface metal atoms are oxidized and dissolved as metal ions into the oxide. On the other hand, dissolved metal ions are reduced and deposited as bulk metal at the interface. These two processes are described by the Butler-Volmer equation \cite{Butler1924,Erdey1930} 
\begin{eqnarray}
\label{eq:butlervolmer}
        r = \frac{k_B T}{h} \Biggl[
        \exp{\left(-\frac{\Delta E_a}{k_B T}\right)} c_\text{metal} \exp{\left(\frac{\left(1-\alpha\right) n e}{k_B T} \Delta \Phi \right)}\nonumber\\
        -
        \exp{\left(-\frac{\Delta E_c}{k_B T}\right)} c_\text{ion} \exp{\left(-\frac{\alpha n e}{k_B T} \Delta \Phi \right)}\Biggr].
\end{eqnarray}
Here, $k_B$ is the Boltzmann constant, $T$ the temperature, and $h$ Planck's constant. The variable $\Delta E_a$ represents the energy barrier for the oxidation and dissolution of bulk metal into metal ions dissolved in the oxide. Reversely, $\Delta E_c$ is the barrier for the reduction and deposition of dissolved metal ions to bulk metal at the electrodes or at the surface of the filament. Accordingly, the first and second summand in Eq.~(\ref{eq:butlervolmer}) are called anodic and cathodic  reaction rate, respectively. Depending on whether the anodic or cathodic rate is larger, the net reaction rate has a positive or negative sign and metal dissolution or deposition is observed. Both $\Delta E_a$ and $\Delta E_c$ critically affect the behavior of $r$. As such, they are derived from \emph{ab initio} calculations, as presented in Sec.~\ref{sec:fluxbarrier}. The remaining variables in Eq.~(\ref{eq:butlervolmer}) are the number $n$ of exchanged electrons in the redox reaction, the electron transfer coefficient $\alpha$, and the elementary charge $e$. The value $\Delta \Phi = \phi_\text{metal} -  \phi_\text{electrolyte}$ denotes the potential drop over the double layer between the metal electrode ($\phi_\text{metal}$) and the solid electrolyte ($\phi_\text{electrolyte}$). The potential $\phi_\text{metal}$ is set to the applied voltage $V$ at interface (1) in Fig.~\ref{fig:fig_model}(b), $\phi_\text{anode}=V$, and to $\phi_\text{cathode}=0$ at interfaces (2) and (3). The potential $\phi_\text{electrolyte}$ is calculated in the electrostatics module, as shown in inset (ii) of Fig.~\ref{fig:workflow}. Finally, $r$ also depends on the reactant and product concentration in the electrode, $c_\text{metal}$, and in the solid electrolyte, $c_\text{ion}$, the latter being time-dependent and passed from the transport module. 

Other important factors that impact the switching characteristics of CBRAM cells are the ion transport properties of the solid electrolyte. They are evaluated in the transport module (inset (iii) of Fig.~\ref{fig:workflow}). The motion of ions through the electrolyte under ion influx/efflux at its interfaces to metal is governed by the following drift-diffusion equation \cite{Strukov2012}
\begin{eqnarray}
\label{eq:ionictransport1}
        \frac{d c_\text{ion}}{dt} + \nabla \cdot J_\text{ion} = 0,
\end{eqnarray}
with
\begin{eqnarray}
\label{eq:ionictransport2}
        J_\text{ion} = - D \nabla c_\text{ion} - n \frac{D}{RT} F c_\text{ion} \nabla \phi_\text{electrolyte}. 
\end{eqnarray}
The first term of the ion flux density, $J_\text{ion}$, is the Fick diffusion flux density, which is driven by the concentration gradient of the ions dissolved in the solid electrolyte. The diffusion coefficient $D$ determines how fast the dissolved cations migrate through the electrolyte. It is expressed as \cite{Ielmini2017}
\begin{eqnarray}
    \label{eq:arrhenius}
        D = D_0 \exp{\left( -\frac{\Delta E_D}{k_B T} \right)}.
\end{eqnarray} 
Both the diffusion coefficient at infinite temperature $D_0$ and the energy barrier for diffusion $\Delta E_D$ can be derived from \emph{ab initio} calculations (Sec.~\ref{sec:diffusion}). The second term in Eq.~(\ref{eq:ionictransport2}) represents the drift flux density driven by the electric field, $\bf{E} = -\nabla \phi_\text{electrolyte}$. Furthermore, $R$ is the molar gas constant and $F$ is the Faraday constant. The transport module is linked to both the electrostatics and the electrochemical modules. On the one hand, $\phi_\text{electrolyte}$ is a direct input to Eq.~(\ref{eq:ionictransport1}). On the other hand, $r$ enters Eq.~(\ref{eq:ionictransport1}) as a boundary condition $J_\text{ion}=r$ at interfaces (1)-(3) in Fig.~\ref{fig:fig_model}(b) and $J_\text{ion}=0$ at the remaining interfaces. 

The electrostatic potential in the solid electrolyte ($\phi_\text{electrolyte}$) is calculated in the electrostatics module (inset (iv) of Fig.~\ref{fig:workflow}) by solving Poisson's equation
\begin{equation}
\label{eq:electrostatics}
    \nabla \cdot \left( \varepsilon_\text{electrolyte} \nabla \phi_\text{electrolyte} \right) = -\varrho.
\end{equation}
Above, $\varepsilon_\text{electrolyte}$ is the relative permittivity of the solid electrolyte. The charge density $\varrho$ is directly proportional to $c_\text{ion}$, which is obtained from the transport module, according to
\begin{equation}
    \varrho = n F c_\text{ion}.
\end{equation}
At the metal/solid electrolyte interfaces, diffusive double layers are formed through the accumulation of ions, which impacts the potential distribution. Due to the nanoscale dimensions of our investigated CBRAM devices, these double layers may overlap \cite{Valov2011}. We therefore treated the metal/solid electrolyte interfaces as a double layer described by a plate capacitance $C_h$. It is included in Eq.~(\ref{eq:electrostatics}) with the surface charge boundary condition $-\textbf{n} \cdot \epsilon_\text{electrolyte} \textbf{E} = C_h (\phi_\text{metal}-\phi_\text{electrolyte})$ at interfaces (1)-(3) in Fig.~\ref{fig:fig_model}(b). At all other interfaces, $-\textbf{n} \cdot \epsilon_\text{electrolyte} \textbf{E} = 0$  applies, where $\bf{n}$ is the unit vector.    

The current flowing through the CBRAM cell upon application of an external voltage is evaluated in the current module (inset (v) of Fig.~\ref{fig:workflow}). The total current density $J_\text{tot}$ is the sum of the electric $J_\text{tu}$ and the ionic $J_\text{ion}$ current density. In the LRS $J_\text{tu}$ is dominated by quantum mechanical tunneling from the filament tip to the closest electrode. Therefore, it can be evaluated with a modified Simmons equation \cite{Menzel2017}
\begin{equation}
\label{eq:tunneling}
    J_\text{tu} = C \frac{\sqrt{2 m_\text{eff} \Delta E_\text{tu}}}{x} \left(\frac{e}{h}\right)^2  \exp{\left( -\frac{4 \pi x}{h} \sqrt{2 m_\text{eff} \Delta E_\text{tu}} \right)} V.  
\end{equation}
In Eq.~(\ref{eq:tunneling}), $C$ is a fitting factor, $m_\text{eff}$ the electron effective mass within the solid electrolyte, $\Delta E_\text{tu}$ the tunneling barrier height, $x$ the tunneling distance, and $V$ the voltage across the tunneling gap. Both $\Delta E_\text{tu}$ and $C$ can be obtained from \emph{ab initio} quantum transport calculations, as detailed in Sec.~\ref{sec:tunneling}. The variable $x$ is time-dependent and is defined by the actual filament shape, as provided from the deformation module. The ionic current density is directly related to $r$ from Eq.~(\ref{eq:butlervolmer}) through
\begin{equation}
    J_\text{ion} = n F r. 
\end{equation}
Above, $n$ is the charge of the ionic species. The total current is then obtained by integrating over the filament surface $A$,
\begin{equation}
\label{eq:totalcurrent}
    I_\text{tot} = \iint \big( J_\text{tu} + J_\text{ion} \big) dA. 
\end{equation}

Joule heating can occur under high electrical current densities. It is the process by which electrons transfer energy to the surrounding lattice, thus exciting crystal vibrations and producing heat. To determine the temperature distribution around the filament resulting from this phenomenon, a Joule heating module must be introduced into the FEM model (inset (vi) of Fig.~\ref{fig:workflow}). It solves the heat equation \cite{Menzel2017}
\begin{equation}\label{eq:heat}
    \rho_m C_p \frac{\partial T}{\partial t} - \nabla \kappa \nabla T = \frac{J_\text{tu}^2}{\sigma}. 
\end{equation}
In Eq.~(\ref{eq:heat}), $\rho_m$ is the mass density of the heated material, $C_p$ its specific heat capacity, $\kappa$ its thermal conductivity, and $\sigma$ its electrical conductivity.  Since $J_\text{tu}$ is evaluated numerically according to Eq.~(\ref{eq:tunneling}), a virtual current source is inserted into the FEM model by placing a current-driven terminal at the tip of the moving filament. The value of the current at this terminal is set to $I_\text{tu}$ in Eq.~(\ref{eq:totalcurrent}). To account for the nanoscale dimensions of the filament, both $\kappa_\text{filament}$ and $\sigma_\text{filament}$ are derived from electrical and thermal quantum transport simulations presented in Sec.~\ref{sec:conductivities}. In contrast, the conductivity of bulk materials are well-established and taken from literature \cite{Kleiner1996,Ho1972,Smith1995,Furukawa1974}. As the black arrows in Fig.~\ref{fig:workflow} indicate, all six modules of the FEM model, i.e.\ Eqs.~(\ref{eq:butlervolmer}), (\ref{eq:ionictransport1}), (\ref{eq:electrostatics}), (\ref{eq:totalcurrent}), and (\ref{eq:heat}), depend on each other. Hence, they must be solved self-consistently. 

Finally, for the sake of completeness, a difference between a traditional electrolysis setup, as  used to derive the Butler-Volmer equation, and our CBRAM model should be mentioned. Amorphous SiO$_2$, which is used as solid electrolyte in our model system, is classified as an untypical solid electrolyte \cite{Menzel2015}. Such solid ionic conductors do not initially contain any metallic cation. Upon dissolution of ions (here Ag$^+$) in the electrolyte, the dielectric layer becomes charged. To maintain charge neutrality and keep the electrochemical reaction going, OH$^-$ counter ions must be introduced from the cathode up to a predefined concentration \cite{Menzel2017}. To summarize, our FEM model features the simulation of the forming step (first SET process) as well as the consecutive full \textit{I-V} hysteretic sweeps, properly accounting for the metal/solid electrolyte interfaces.

\subsection{Modules to Extract the Material Parameters}

\subsubsection{Evaluation of the Ionic Diffusion Parameters}\label{sec:diffusion}

The ionic transport properties of metallic ions in a solid-state electrolyte largely depend on their diffusion parameters. To model ionic transport in oxides from first-principles, we have developed a DFT-based module that provides the diffusion coefficient $D$ of metals in amorphous insulators by combining the Einstein-Smoluchovski relation and the Arrhenius equation. The Einstein-Smoluchovski relation connects the mobility of a diffusing particle with the square of the mean free path $\big \langle r^2 \big \rangle$ and the time $t$ through \cite{Sahli2007}
\begin{equation}
\label{eq:diffusion}
	D = \frac{\big \langle r^2 \big \rangle}{6t}.
\end{equation}
The influence of the temperature is described through Arrhenius' Eq.~(\ref{eq:arrhenius}) so that the diffusion constant at infinite temperature $D_0$ and the energy barrier for diffusion $\Delta E_D$ are the only parameters to be computed \cite{Sahli2007,Sankaran2012}. 

All DFT calculations are carried out with the CP2K package \cite{Kuhne2020} with contracted Gaussian-type, double-zeta valence polarized (DZVP) orbitals as basis set \cite{VandeVondele2007}, the Perdew-Burke-Ernzerhof (PBE) exchange-correlation functional \cite{Perdew1996},  Goedecker-Teter-Hutter (GTH) pseudopotentials \cite{Goedecker1996}, and periodic boundary conditions along all three dimensions. No $k$-point grid is used; all calculations are performed at the $\Gamma$-point. Geometry relaxations are performed using the Broyden-Fletcher-Goldfarb-Shanno (BFGS) \cite{Fletcher1970} optimizer with a convergence criteria of $4.5 \cdot 10^{-4}$~Ha/Bohr for forces, $3\cdot 10^{-3}$~Bohr for the geometry variations, $3 \cdot 10^{-4}$~Ha/Bohr for the root-mean-square (RMS) forces, and $1.5\cdot 10^{-3}$~Bohr for the RMS geometry variations. \emph{Ab initio} molecular dynamics (AIMD) simulations are carried out in the NVT ensemble using the canonical sampling through velocity rescaling (CSVR) thermostat \cite{Bussi2007} with a time step of 1~fs. The band of NEB calculations is split into five to ten replica so that two adjacent replica are located ca.\ 1~\AA\ apart from each other. The direct inversion in the iterative subspace (DIIS) \cite{Pulay1980} algorithm is applied for the band optimization with the convergence criteria as before for the forces, but $2\cdot 10^{-3}$ ($10^{-3}$) Bohr for the (RMS) geometry variations. 

\begin{figure}
\includegraphics[width=7.5cm]{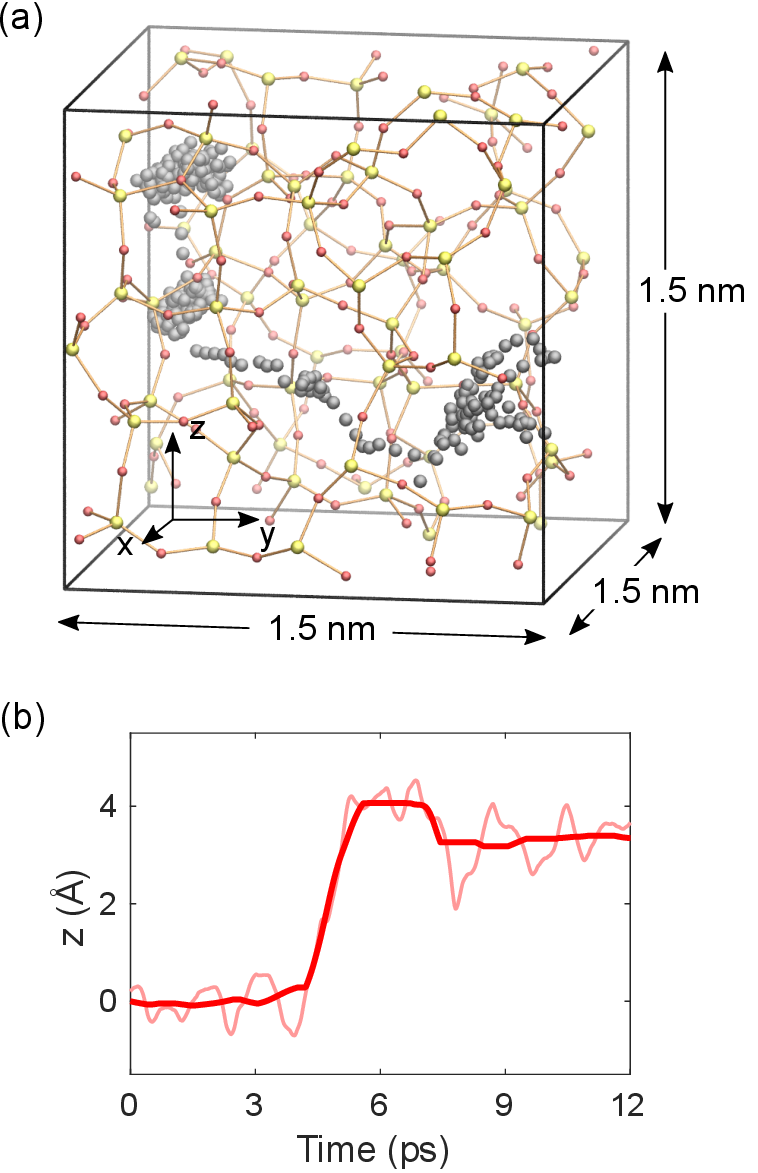}
\caption{\label{fig:fig_difftraj} (a) Diffusion trajectory of an Ag$^+$ ion through a cube of a-SiO$_2$ with a side length of 1.5 nm at 1200~K. The position of the Ag$^+$ ion, displayed as gray spheres, is shown every 0.1~ps for a duration of 39~ps. The yellow and red spheres represent Si and O atoms, respectively. (b) $z$ component of a 12~ps long inset of the diffusion trajectory in (a) before (thin line) and after (thick line) the application of the smoothing algorithm of Ref.~\cite{Sahli2007}. }
\end{figure}

To calculate $D_0$ and $\Delta E_D$ in Eq.~(\ref{eq:arrhenius}), cubes of amorphous oxides with a side length of 1.5 nm are generated by the melt-and-quench approach \cite{Ducry2020} with classical force-field \cite{Pedone2006} molecular dynamics (MD), as implemented in the LAMMPS \cite{Plimpton1995} tool. Specifically, oxides with a predefined density and volume are melted at 4000~K for several nanoseconds before the melt is quenched to 300~K with a cooling rate of 15~K/ps using MD in the NVT ensemble with a time step of 1~fs. Then, the samples of force-field oxides are annealed with DFT to equilibrate the obtained atomic structures. Only samples without coordination defects are chosen for the subsequent diffusion studies. To obtain statistics, several configurations are selected and a single charged metallic ion interstitial is inserted into each sample. Next, AIMD is performed for at least 35~ps at different temperatures. For our Ag/a-SiO$_2$/Pt CBRAM stack, eleven different a-SiO$_2$ samples with a density of 2.20~g/cm$^3$ were examined. Diffusion trajectories were sampled at six different temperatures (700~K, 800~K, 900~K, 1000~K, 1200~K, and 1500~K) for a duration of ca.\ 40~ps. An in-depth analysis of the created diffusion trajectories, as shown in Fig.~\ref{fig:fig_difftraj}(a) for an Ag$^+$ cation in a-SiO$_2$, revealed that the metal ion oscillates around equilibrium centers in pores during most of the time. Transitions between pores are recorded only a few times per sample and typically last less than a picosecond. 

Before $\big \langle r^2 \big \rangle$ can be calculated and inserted into Eq.~(\ref{eq:diffusion}), thermal noise in the ion diffusion trajectories is eliminated by applying a smoothing procedure \cite{Sahli2007}. It consists of two main steps. First, a moving average over $2n_a+1$ time steps is applied. The resulting positions are calculated according to
\begin{eqnarray}
    x_i' && = \text{mean} \big( x_{i-n_a}, x_{i-n_a+1}, ..., x_{i+n_a-1}, x_{i+n_a} \big) \nonumber \\
    &&= \frac{1}{2n_a+1} \sum_{k=i-n_a}^{i+n_a} x_k.
\end{eqnarray}
In the second step, the positions of the final smoothed trajectory are constructed through the application of a median filter with a symmetric window of $2n_m+1$ time steps,  
\begin{equation}
    x_i'' = \text{median} \big( x_{i-n_m}', x_{i-n_m+1}', ..., x_{i+n_m-1}', x_{i+m_m}' \big). 
\end{equation}
The time ranges $n_a$ and $n_m$ must be chosen such that thermal oscillations are removed as much as possible without blurring transitions between oscillation centers. For Ag$^+$ ions in a-SiO$_2$ we found that $n_a=500$ and $n_m=1000$ (with a time step of 1~fs in the MD trajectory) satisfy these criteria most suitably.  Exemplary results of this procedure are shown in Fig.~\ref{fig:fig_difftraj}(b). 

\begin{figure}
\includegraphics[width=8.6cm]{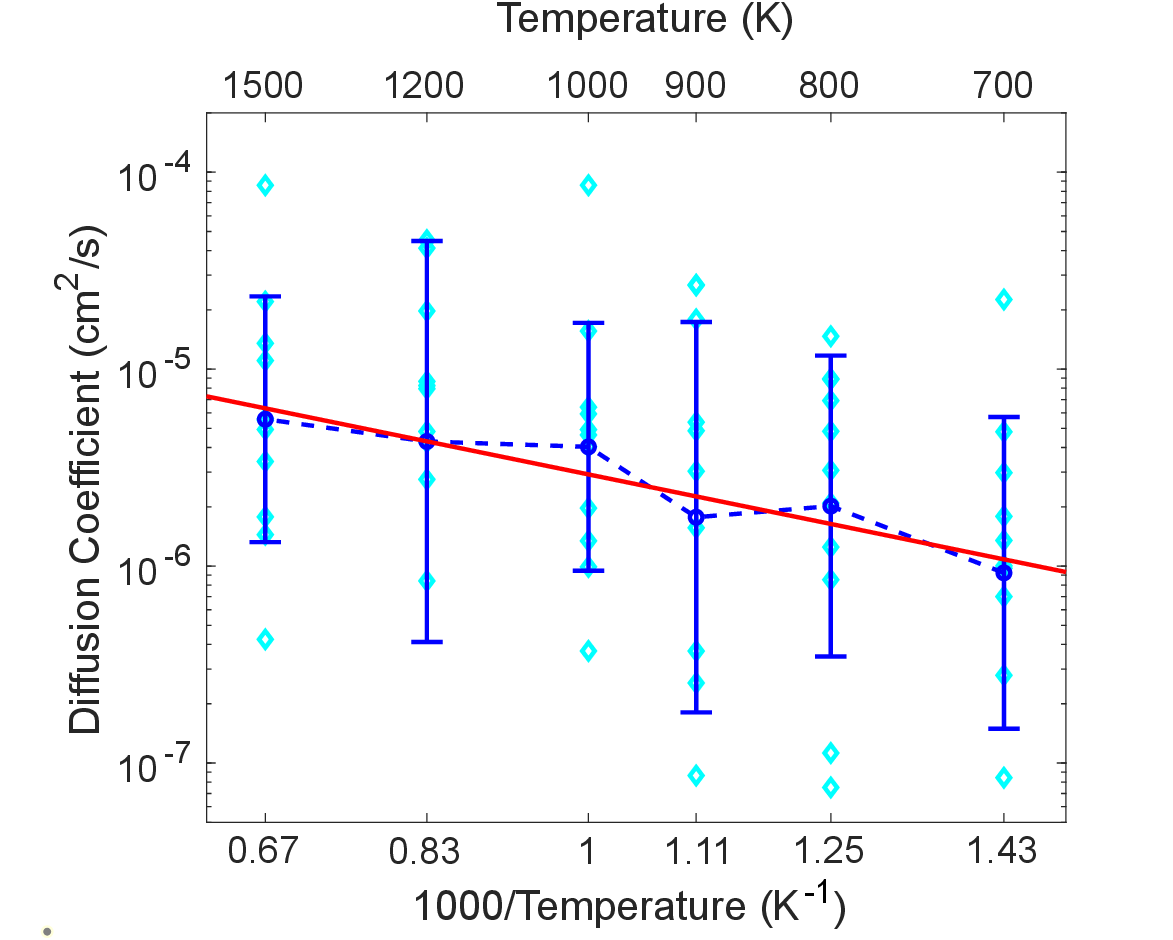}
\caption{\label{fig:fig_arrh} Arrhenius plot of the diffusion coefficient $D$ of Ag$^+$ cations dissolved in a-SiO$_2$ vs.\ the reciprocal temperature based on 25~ps long trajectories. The light blue dots represent individual measurements of the diffusion coefficient coming from eleven different samples, the dashed blue line is their average at each temperature, the error bars denote the standard deviation, and the red line is a linear fit of the mean values. }
\end{figure}

\begin{figure*}
\includegraphics[width=17.6cm]{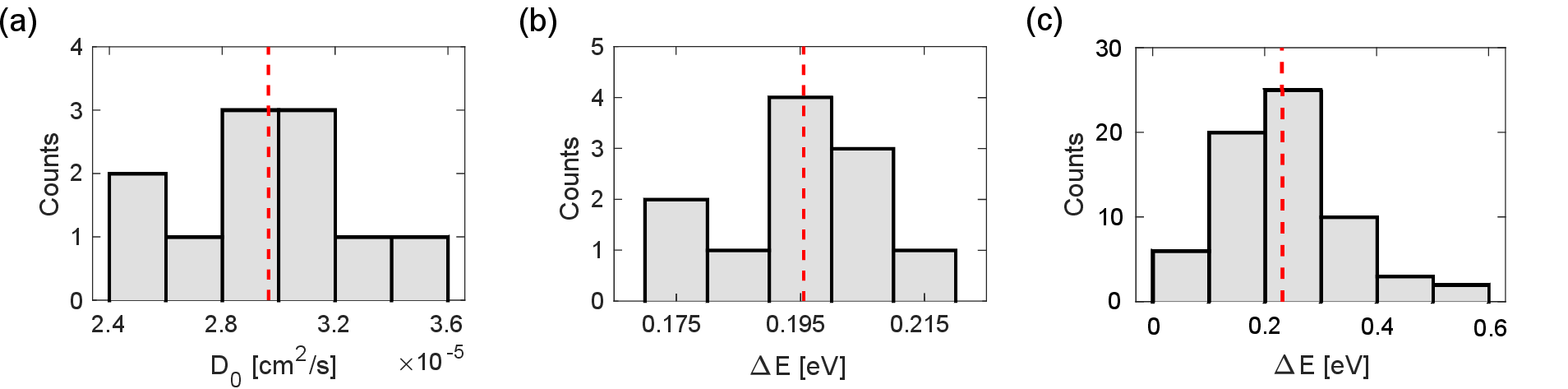}
\caption{\label{fig:fig_histo} Histogram of (a) the diffusion prefactors $D_0$ and (b) energy barriers $\Delta E_D$ extracted from Arrhenius plots based on diffusion trajectories evaluated at time lengths ranging from 18 to 28~ps for Ag$^+$ cations in a-SiO$_2$. The dotted red line represents the mean value. (c) Histogram of energy barriers derived via NEB calculations based on all Ag$^+$ ion transitions recorded by AIMD.}
\end{figure*}

An Arrhenius plot, as displayed in Fig.~\ref{fig:fig_arrh}, can be created with the smoothed ion trajectories. The first 5~ps of each AIMD trajectory is discarded for equilibration. Then, the square mean free path is evaluated by calculating the distance between the start and end point of the diffusion trajectory for each sample. Using Eq.~(\ref{eq:diffusion}) the diffusion coefficient $D$ can be determined for each sample and temperature. Figure~\ref{fig:fig_arrh} indicates that the obtained diffusion coefficients can significantly differ between different samples, but their average follows the expected Arrhenius-type behavior predicted by Eq.~(\ref{eq:arrhenius}). To avoid that $\big \langle r^2 \big \rangle$ depends on the chosen time length $t_i$ of the considered diffusion trajectory, the described procedure is repeated for various $t_i \in \big[ 18, 19, .., 27, 28 \big]$~ps for the aforementioned trajectories of Ag$^+$ through a-SiO$_2$. Based on the averaged diffusion coefficients over all evaluated $t_i$, the parameters $D_0$ (Fig.~\ref{fig:fig_histo}(a)) and $\Delta E_D$ (Fig.~\ref{fig:fig_histo}(b)) can be extracted. For an Ag$^+$ cation in a-SiO$_2$, $\Delta E_D = 0.196 \pm 0.013$~eV and $D_0 = 2.962 \cdot 10^{-5} \pm 2.907 \cdot 10^{-6}$~cm$^2$/s were found. 

To validate the chosen approach, the value of $\Delta E_D$ obtained for the Ag$^+$ cations dissolved in a-SiO$_2$ was confirmed by performing NEB calculations. All ionic transitions between oscillation centers recorded during AIMD were collected, leading to a total of 33 different transitions. The corresponding equilibrium center of each pore was evaluated through atomic relaxation upon fixed lattice cells and taken as start and end point of the NEB trajectories. A NEB chain is formed of six to ten replica, depending on the distance between the start and end point. The energy barrier height of both the forth and back reaction of the NEB trajectories were then calculated using climbing-image-NEB \cite{Henkelman2000cineb}. Figure~\ref{fig:fig_histo}(c) shows the histogram for all 66 investigated transition barriers. An average diffusion barrier of $0.234 \pm 0.122$~eV is obtained, close to the 0.196~eV determined with AIMD, but with a higher uncertainty.

\subsubsection{Determination of the Tunneling Current Parameters}\label{sec:tunneling}

\begin{figure}
\includegraphics[width=8.6cm]{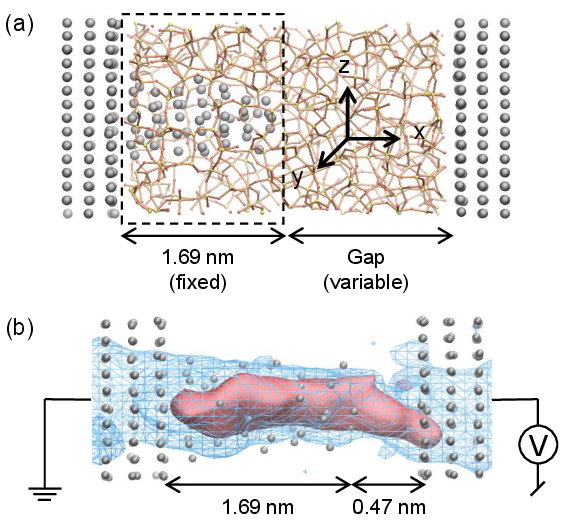}
\caption{\label{fig:fig_tunneling} (a) Schematic view of the 3-D atomic CBRAM structure consisting of two $\langle$111$\rangle$-oriented Ag plates surrounding a chunk of a-SiO$_2$ used in quantum transport calculations. The gray spheres represent Ag atoms, the orange network the a-SiO$_2$ matrix. The cross section along the \emph{y} and \emph{z} axes (assumed periodic) measures $2.06 \times 2.04$~nm$^2$. A 1.69~nm long, crystalline Ag filament is inserted into the a-SiO$_2$ matrix. The left part with the filament (marked with the dashed black box) remains the same, regardless of the gap length. By varying the length of the gap from 0 to 1.69~nm, different conductance states can be mimicked. (b) Spatial distribution of the ballistic current flowing along the \emph{x} axis in the structure of (a) in an intermediate conductance state with a gap of 0.47~nm. An external voltage $V = 1$~mV is applied. The current is displayed as two different isosurfaces (solid and wireframe). }
\end{figure}

The electronic current flowing through CBRAM cells is evaluated by fitting tunneling current results obtained by \emph{ab initio} quantum transport calculations to the modified Simmons Eq.~(\ref{eq:tunneling}). To determine $\Delta E_\text{tu}$ and $C$ in this expression, atomistic filamentary CBRAM structures with varying oxide lengths are first generated by combining DFT and MD. To begin with, two metal electrodes are attached to a layer of amorphous oxide obtained as described in Sec.~\ref{sec:diffusion}, resulting in metal/oxide/metal systems of more than 3000 atoms. The amorphous oxide is free from coordination defects to avoid a defect-induced increase of the electrical conductance by, e.g., trap assisted tunneling. Then, a cylinder with a cross section $A_\text{filament}$ and a length $l_\text{filament}$ is cut out of the left part of the oxide and replaced by a size-equivalent, crystalline nano-filament of the same metal atoms as the electrodes. This results in structures as the one shown in Fig.~\ref{fig:fig_tunneling}(a) for an Ag/a-SiO$_2$/Pt CBRAM cell. Note that for computational convenience, two Ag electrodes are used instead of an Ag and Pt one. This simplification has a negligible impact on the current magnitude \cite{Ducry2018}. To obtain a set of CBRAM cells with different oxide gaps separating the filament tip and its closest electrode only one single sample of amorphous oxide is needed. The oxide is tailored to the desired length by cutting away parts of it on the right-hand side while the left-hand side with the embedded filament remains untouched. This allows to keep the shape of the filament and hence its interfaces with the oxide the same for all samples. Due to the varying oxide lengths of the cells, the gap between the tip of the filament and the electrode varies, mimicking different filament growth stages. Finally, each structure is optimized and annealed for several picoseconds with AIMD to reduce the overall strain and minimize its total energy. 

As next step, the Hamiltonian $H$ and overlap matrices $S$ of the atomic configurations are calculated with CP2K. While it is important to relax atomic geometries with a DZVP basis set, SZV is generally sufficient to evaluate the transport properties around the Fermi energy for the atomic species present in this work \cite{Ducry2017}. The $H$ and $S$ matrices are then passed to the OMEN quantum transport solver \cite{Luisier2006}, which solves the Schr\"odinger equation with open boundary conditions using \emph{ab initio} inputs. The transmission function $T_e(E)$ from one electrode to the other is computed in this scheme with an energy resolution of typically $dE=1$~meV. By applying the Landauer-B\"uttiker formula \cite{Buttiker1985}
\begin{equation}
    \label{eq:landauer}
    I_{d,\Delta V} = -\frac{e}{\hbar} \int \frac{\text{d}E}{2\pi} T_e(E) \bigl( f(E,E_{F,L},T)-f(E,E_{F,R},T) \bigr),
\end{equation}
the ballistic electrical current can be obtained. In Eq.~(\ref{eq:landauer}), $f(E,E_{F,L},T)$ ($f(E,E_{F,R},T)$) is the Fermi-Dirac distribution function of electrons situated in the left (right) contact with energy $E$, Fermi energy $E_{F,L}$ ($E_{F,R}$), and temperature $T$. Finally, $\hbar$ is Planck's reduced constant. Note that $E_{F,R}=E_{F,L}-e \Delta V$, $\Delta V=1$~mV being the applied voltage. In this approach we did not solve the Poisson equation as the applied voltage in the order of millivolt is not large enough to significantly modify the electrostatics obtained in CP2K at equilibrium. 

\begin{figure}
\includegraphics[width=8.6cm]{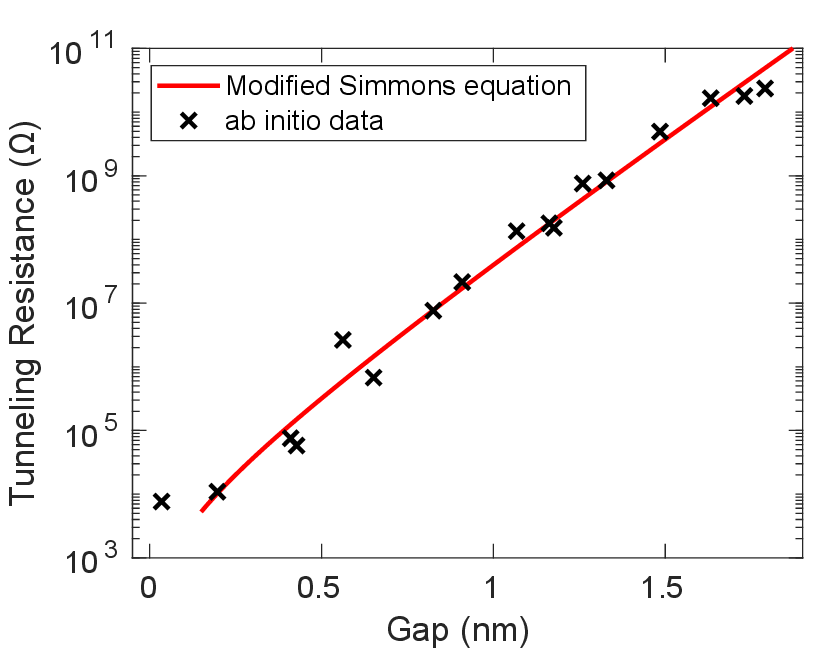}
\caption{\label{fig:fig_simmon} Tunneling resistance computed with an \emph{ab initio} quantum transport approach for the Ag/a-SiO$_2$/Ag CBRAM structures of Fig.~\ref{fig:fig_tunneling}(a). A total of 17 different samples with varying oxide gap lengths ranging from  0 to 1.69~nm are reported. The \emph{ab initio} data (black crosses) is fitted to the modified Simmons Eq.~(\ref{eq:tunneling}) (red line) with the electronic parameters $C = 5.1$, $\Delta E_\text{tu} = 1.03$~eV, and $m_\text{eff} = 0.63 m_0$. }
\end{figure}

In OMEN, atomically-resolved current trajectories can be produced. An exemplary current map is reported in Fig.~\ref{fig:fig_tunneling}(b). As expected, the current density is highest in the filament region, but a non-negligible tunneling current density is visible in the oxide. The electrical conductance $G_e$ can then be derived from the expression
\begin{equation}
    \label{eq:elconductance}
    G_e = \frac{d I_{d, \Delta V}}{d V}.
\end{equation}
The obtained results from the quantum transport calculations can eventually be used to determine $C$ and $\Delta E_\text{tu}$ in Eq.~(\ref{eq:tunneling}) through fitting. To do so, the tunneling resistance, which is the inverse of $G_e$, is plotted against the filament gap length in Fig.~\ref{fig:fig_simmon}. It can be seen there that the results from 17 different Ag/a-SiO$_2$/Ag CBRAM cells, with gap lengths varying between 0~nm and 1.67~nm, can be very well fitted by Eq.~(\ref{eq:tunneling}). The solid red line in Fig.~\ref{fig:fig_simmon} was obtained by setting $C = 5.1$, $\Delta E_\text{tu} = 1.03$~eV, and $m_\text{eff} = 0.63 m_0$ (fixed to the experimental value \cite{Wen1998}, with $m_0$ being the electron's mass) in Eq.~(\ref{eq:tunneling}). To convert the tunneling current $I_{d,\Delta V}$ to the tunneling current density $J_\text{tu}$, a filament tip cross section of $A_\text{filament} = 0.76$~nm$^2$ was taken. 

\subsubsection{Evaluation of the Reaction Barriers at the Metal/Oxide Interfaces} \label{sec:fluxbarrier}

At the metal/solid electrolyte interfaces, redox reactions take place according to Eq.~(\ref{eq:butlervolmer}). Two parameters that critically affect the behavior of $r$ in Eq.~(\ref{eq:butlervolmer}) are $\Delta E_a$ and $\Delta E_c$. To recall, $\Delta E_a$ is the energy barrier for the oxidation and dissolution of bulk metal into metal ions dissolved in the oxide. Reversely, $\Delta E_c$ is the barrier for the reduction and deposition of dissolved metal ions to bulk metal at the electrodes or at the surface of the filament. In this section, a procedure to derive $\Delta E_a$ and $\Delta E_c$ from \emph{ab initio} calculations is presented.  

\begin{figure*}
\includegraphics[width=17.6cm]{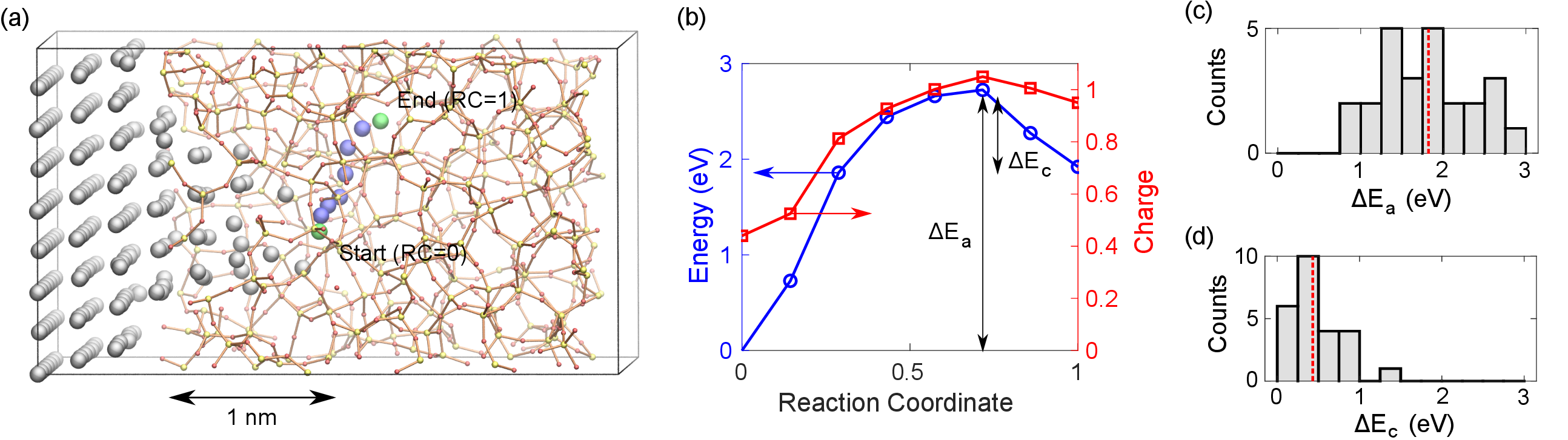}
\caption{\label{fig:fig_iongeneration} (a) Atomistic illustration of the nudged elastic band method used to extract $\Delta E_a$ and $\Delta E_c$. The constructed structure comprises an Ag electrode with an attached filament (left, gray spheres) surrounded by an a-SiO$_2$ matrix (orange network). The bottom green Ag atom at the filament tip is taken as starting point for the NEB calculation (reaction coordinate (RC) 0). The upper green sphere represents the same Ag$^+$ ion now residing in a pore after dissolution. It is taken as end point (reaction coordinate 1). The blue spheres depict the positions of the Ag$^+$ ions in the intermediate replica. (b) Evolution of the energy (blue line with circles) and Mulliken charge (red line with squares) of the diffusing Ag$^+$ cation as a function of the reaction coordinate (RC) of the NEB calculation shown in (a). The collected barrier heights (c)  $\Delta E_a$ and (d) $\Delta E_c$ are summarized as histograms containing the results from 25 samples. The dashed red lines represent the mean values of $\Delta E_a$ and $\Delta E_c$. }
\end{figure*}

For the metal atoms from the bulk electrodes or from the nano-filament seed to be dissolved in the oxide, an energy barrier $\Delta E_a$ must be overcome, as depicted in Fig.~\ref{fig:fig_iongeneration}(a)-(b). There are several factors contributing to the shape of this energy barrier: (i) The transition from bulk metal to a metal cation is associated with a charge transfer, which depends on the energy barrier of the corresponding redox reaction. (ii) Cohesion energy is lost when an ion leaves the bulk metal. (iii) The newly dissolved metal cation and the surrounding oxide matrix atoms may interact with each other and even form bonds, thus contributing to the total energy barrier. This also includes diffusion barriers as the cation migrates through the oxide. (iv) Once cations have overcome the maximum barrier height, we have found that they typically settle into a pore of the host oxide. This process releases energy in form of solvation and electrochemical bonding energies with nearby atoms. In the reverse case, when a solvated metal cation is deposited back onto a bulk metal interface, an energy $\Delta E_c$ is needed to overcome the reaction barrier. Typically, $\Delta E_c$ is lower than $\Delta E_a$. The aforementioned four contributions to the energy barriers can significantly differ in solid-state electrolytes, as compared to well-studied electrolysis systems with liquid electrolytes. Owing to the lack of dedicated experimental data it is necessary to determine $\Delta E_a$ and $\Delta E_c$ from first-principles. To this end we employ DFT and the nudged elastic band method. 

To study the reaction barriers in an Ag/a-SiO$_2$/Pt device, 25 different structures of amorphous SiO$_2$ free from coordination defects were created with the melt-and-quench approach from Sec.~\ref{sec:diffusion}. A cave was carved out from these oxide matrices such that an Ag filament containing 37 to 49 Ag atoms can be inserted. This allows to consider filaments with different sizes and shapes. Before the filament was introduced, the oxide interface was smoothed by reducing the number of dangling Si and O bonds through manual adjustment of the interface atoms and multiple annealing steps with AIMD at 600~K. With this procedure, more realistic Ag/a-SiO$_2$ interfaces can be realized, as the filament grows through a chain of pores. After inserting the Ag filament into the oxide, two $\langle$111$\rangle$-oriented bulk Ag electrodes with a cross section of $2.06 \times 2.04$~nm$^2$ were attached to both extremities of the oxide layer. The resulting structure was annealed again with AIMD during a short time and optimized. It is then ready for subsequent NEB calculations, as shown in Fig.~\ref{fig:fig_iongeneration}(a). 

To model the transfer of a bulk Ag atom from the filament tip into the surrounding oxide, which gives rise to an Ag$^+$ ion, appropriate starting and end points must be chosen and passed to the nudged elastic band method. As solvated Ag$^+$ ions typically reside in pores of the host oxide at equilibrium, a structural analysis of the a-SiO$_2$ is first performed. By defining a pore as a region free from Si or O atom neighbors in a range of 3~\AA, it was found that each sample contains two to nine pores. They were defined as potential end points for the NEB. After evaluating potential migration paths between these pore sites and Ag surface atoms on the filament tip, the most promising combination was selected as starting and end points for the NEB. For each starting/end point, improved-tangent-NEB \cite{Henkelman2000itneb} calculations were performed with at least five intermediate replicas. An exemplary NEB trajectory is displayed in Fig.~\ref{fig:fig_iongeneration}(a). Its corresponding energy and Mulliken charge evolution in Fig.~\ref{fig:fig_iongeneration}(b) reveals that that the already slightly positively charged Ag tip atom is fully oxidized upon solvation in the oxide. 

The collected energy barriers $\Delta E_a$ and $\Delta E_c$ are displayed as histograms in Figs.~\ref{fig:fig_iongeneration}(c) and \ref{fig:fig_iongeneration}(d), respectively. A mean value $\Delta E_a = 1.80 \pm 0.57$~eV and $\Delta E_c = 0.44 \pm 0.30$~eV are extracted. As discussed in Sec.~\ref{sec:result_iv}, $\Delta E_a$ and $\Delta E_c$ must be slightly adjusted in the FEM-based CBRAM model ($\Delta E_a = 1.85$~eV and $\Delta E_c = 0.49$~eV) to ensure numerical stability. However, it is worth noting that the adjusted values still clearly lie within the standard deviation of our DFT results.

\subsubsection{Determination of the Electrical and Thermal Conductivity of the Filament} \label{sec:conductivities}

The extent of Joule heating in the CBRAM cell is modeled by Eq.~(\ref{eq:heat}). Besides the geometry of the filament embedded in the solid-state electrolyte, its electrical ($\sigma$) and thermal ($\kappa$) conductivity are critical material parameters. Because of the nanoscale dimensions of this filament bulk values are expected to be significantly altered \cite{Ducry2017}. Hence, these parameters should be directly evaluated in dedicated atomic systems. To create them, Ag/a-SiO$_2$/Pt CBRAM cells were constructed as in Sec.~\ref{sec:tunneling}. Ag electrodes with a cross section of $2.06 \times 2.04$~nm$^2$ were placed on the right and left side of an a-SiO$_2$ layer. 

\begin{figure}
\includegraphics[width=8.6cm]{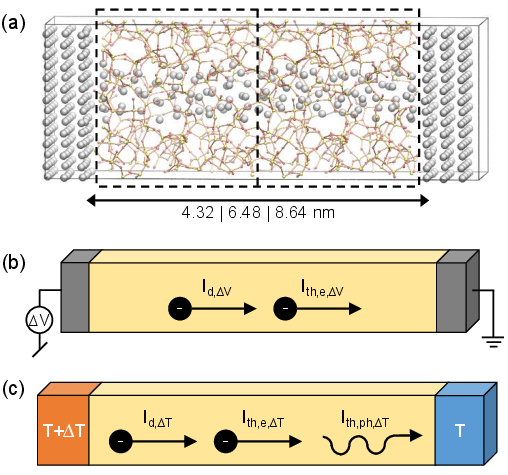}
\caption{\label{fig:fig_cond_theory} (a) CBRAM structure consisting of two Ag bulk electrodes and repetitions of 2.16~nm long a-SiO$_2$ blocks (dashed rectangular boxes). A bridging Ag filament with a diameter of 7~\AA\ is inserted in between. The a-SiO$_2$ blocks are repeated two, three, and four times resulting in oxide lengths of 4.32, 6.48, and 8.64~nm. (b) Electrical ($I_{d,\Delta V}$) and thermal ($I_{th,e,\Delta V}$) currents flowing when a voltage $\Delta V$ is applied to the left electrode of the CBRAM cell in (a). (c) Electrical and thermal current contributions when a temperature difference $\Delta T$ is applied between the left ($T+\Delta T$) and right ($T$) electrodes. An electrical ($I_{d,\Delta T}$) and a thermal ($I_{th,e,\Delta T}$) current carried by electrons are present. They are accompanied by a second thermal current carried by phonons ($I_{th,ph,\Delta T}$). }
\end{figure}

The electrical or thermal conductivity of samples can be conveniently calculated with the "\emph{dR/dL}" method \cite{Berger1972}, which corresponds to the well-known transmission line measurements (TLM) in experiments. It consists of evaluating the dependence of the electrical or thermal resistance of the considered device with respect to its length. To establish this "resistance vs.\ length" relationship, a-SiO$_2$ blocks measuring 2.16~nm in length are repeated two, three, and four times along the axis joining both metallic electrodes. Hence, the Ag/a-SiO$_2$ interface remains the same regardless of the device length. Next, following the same procedure as in Sec.~\ref{sec:fluxbarrier}, a cylindrical Ag filament with a diameter of 7 \AA\ is inserted into the a-SiO$_2$ matrix such that it bridges both electrodes. Eventually, the entire device structure is optimized at the DFT level with CP2K, resulting in CBRAM cells as shown in Fig.~\ref{fig:fig_cond_theory}(a). When a voltage is applied to such devices, an electrical ($I_{d,\Delta V}$) and a thermal ($I_{th,e,\Delta V}$) current start flowing, as illustrated in Fig.~\ref{fig:fig_cond_theory}(b). When instead a temperature gradient is applied, an electrical ($I_{d,\Delta T}$) and a thermal ($I_{th,e,\Delta T}$) current carried by electrons are present. They are accompanied by a second thermal current carried by phonons ($I_{th,ph,\Delta T}$) (Fig.~\ref{fig:fig_cond_theory}(c)). These five different current types serve as inputs to evaluate the electrical and thermal conductivity of the CBRAM of interest. 

\begin{figure}
\includegraphics[width=8.6cm]{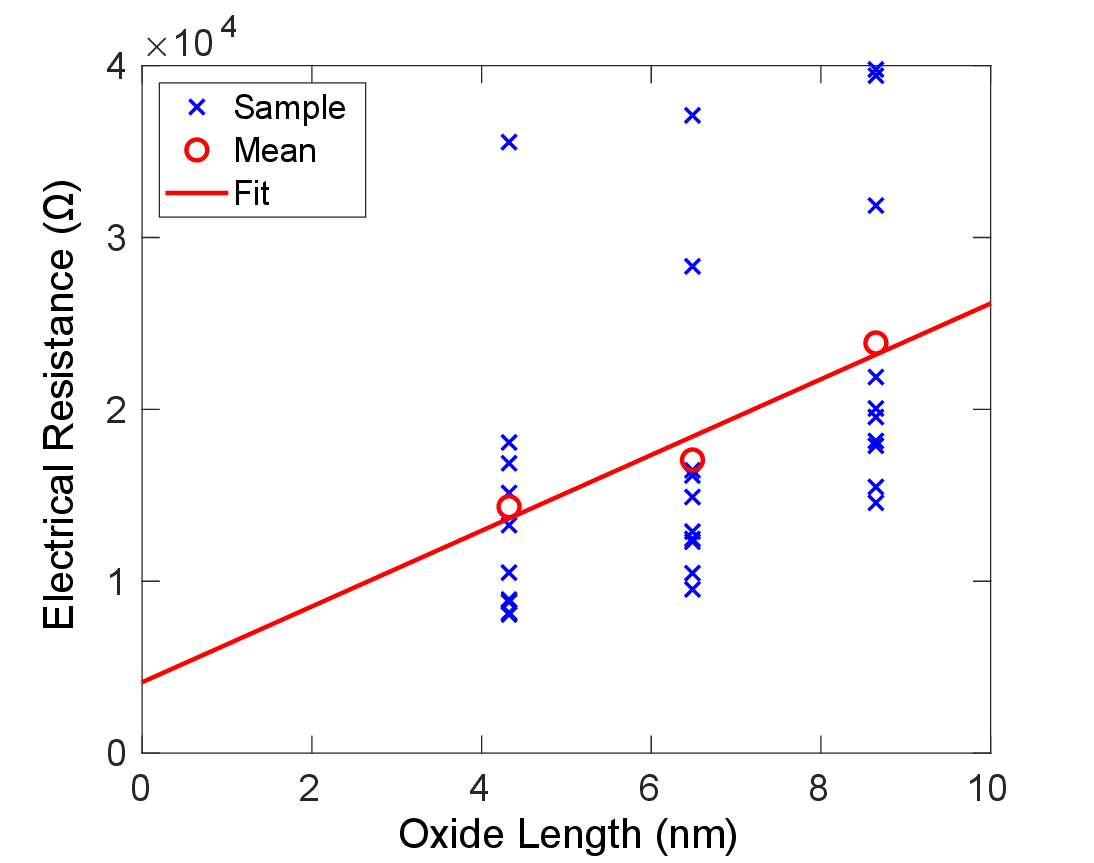}
\caption{\label{fig:fig_elcond} Electrical resistance of the CBRAM cell in Fig.~\ref{fig:fig_cond_theory}(a) vs.\ its oxide length. Ten Ag/a-SiO$_2$/Ag samples with three different oxide lengths each were evaluated. The crosses represent the resistance of each individual sample, the circles are the mean values at each length, and the solid line is a linear fit through the three circles. }
\end{figure}

To calculate the electrical conductivity of the embedded Ag filament, the Hamiltonian ($H$) and overlap ($S$) matrices of the CBRAM cells of different lengths are first created with CP2K using the same DFT settings as in Sec.~\ref{sec:tunneling}. These matrices are then passed to OMEN, which produces the electrical transmission function $T_e(E)$ with an energy resolution of 1~meV. With this quantity, the electrical current $I_{d,\Delta V}$ can be computed with the Landauer-B\"uttiker formula (\ref{eq:landauer}), which is given by  
\begin{equation} \label{eq:landauerlinear}
    I_{d,\Delta V} \approx -\frac{e}{\hbar} \int \frac{dE}{2 \pi} T_e(E) \biggl(\frac{df(E,E_F,T)}{dE} \biggr) \Delta V 
\end{equation}
in the linear approximation, i.e.\ when a small voltage $\Delta V$ is applied. In this equation, $f=f(E,E_F,T)$ is the Fermi-Dirac distribution function of electrons with energy $E$, Fermi energy $E_F$, and temperature $T$. The electrical conductance $G_e$ can then be derived according to Eq.~(\ref{eq:elconductance}). All quantum transport calculations were performed in the coherent limit because scattering is mainly caused by the disorder of the a-SiO$_2$ layer and the inclusion of electron-phonon scattering has almost no influence on the value of $G_e$. To account for the stochastic variability of the amorphous oxide layers, ten different samples were created for each device length. The average over the corresponding electrical resistance $R_e = 1/G_e$ is shown in  Fig.~\ref{fig:fig_elcond}. As expected for diffusive transport, $R_e$ exhibits a linear dependence on the device length $L$
\begin{eqnarray}
        R_e(L) = R_0 + \varrho_\text{filament}L. 
\end{eqnarray}
Above, $R_0$ is the contact and ballistic resistance and $\varrho_\text{filament}$ the resistivity of the filament embedded within an a-SiO$_2$ layer. Figure~\ref{fig:fig_elcond} confirms the validity of the \emph{"dR/dL"} method for our structures. As the electrical conductivity $\sigma$ is the inverse of the resistivity, a value of $\sigma_\text{filament} = 1.18 \cdot 10^6$~S/m is found for the filament in Fig.~\ref{fig:fig_cond_theory}(a). This value is 50 times smaller than in bulk Ag ($\sigma_\text{Ag, bulk} = 6.14 \cdot 10^7$~S/m \cite{Matula1979}). 

The determination of the thermal conductivity $\kappa$ is more complex since two factors contribute to $\kappa$: heat can be carried by electrons and by lattice vibrations (phonons). Hence, $\kappa$ is made of two components, $\kappa_e$ and $\kappa_{ph}$, the former referring to the electronic contribution, the latter to the phonon one. The electronic contribution typically dominates in metals, while phonons carry more heat in semiconductors and oxides. Our metallic filaments being embedded in an amorphous oxide, it is not clear which transport mechanism dominates. Both should therefore be evaluated. As for electrons, the \emph{"dR/dL"} method is applied in both cases. We first derive expressions for the thermal conductance, demonstrate that the corresponding resistance linearly increases with the device length, and finally extract $\kappa_e$ and $\kappa_{ph}$. The electronic component of the thermal conductance, $G_{th,e}$, is defined as \cite{Sivan1986}
\begin{eqnarray}
        G_{th,e} = \frac{dI_{th,e,\Delta T}}{dT} + \bigg( \frac{dI_{th,e,\Delta V}}{dV} \bigg) S,
\end{eqnarray}
where $S$ is the Seebeck coefficient. Thermal current is carried by electrons when a voltage $\Delta V$ ($I_{th,e,\Delta V}$) or when a temperature gradient $\Delta T$ ($I_{th,e,\Delta T}$) is applied. These currents can be calculated from the transmission function $T_e(E)$ as
\begin{equation}
    I_{th,e,\Delta T} = -\frac{e}{\hbar} \int \frac{dE}{2 \pi} (E-E_F) T_e(E) \biggl(\frac{df}{dT} \biggr) \Delta T
\end{equation}
and
\begin{equation}
    I_{th,e,\Delta V} = -\frac{e}{\hbar} \int \frac{dE}{2 \pi} (E-E_F) T_e(E) \biggl(\frac{df}{dE} \biggr) \Delta V.
\end{equation}
In both equations, $E_F$ is the Fermi energy of the source contact. The Seebeck coefficient $S$ is given by
\begin{eqnarray}
        S = \bigg(\frac{dI_{d,\Delta V}}{dV}\bigg)^{-1} \bigg(\frac{dI_{d,\Delta T}}{dT}\bigg).
\end{eqnarray}
Here, $I_{d,\Delta T}$ is the electrical current flowing when there is a temperature difference of $\Delta T$ between the two metallic electrodes. It can be computed with the Landauer-B\"uttiker formula (\ref{eq:landauerlinear}) by setting the same Fermi level to both electrodes, but two different temperatures, $T$ and $T+\Delta T$, so that
\begin{equation} 
    I_{d,\Delta T} = -\frac{e}{\hbar} \int \frac{dE}{2 \pi} T_e(E) \biggl(\frac{df(E,E_F,T)}{dT} \biggr) \Delta T. 
\end{equation}

Finally, the thermal current carried by phonons, $I_{th,ph,\Delta T}$, induces the following conductance 
\begin{eqnarray}
        G_{th,ph} = \frac{dI_{th,ph,\Delta T}}{dT}.
\end{eqnarray}
To compute $I_{th,ph,\Delta T}$, the phonon transmission function $T_{ph}(\omega)$ is needed so that the Landauer-B\"uttiker formula can be utilized in this case too 
\begin{equation}
    I_{th,ph,\Delta T} = \int \frac{d \omega}{2 \pi} \hbar \omega  T_{ph}(\omega) \biggl(\frac{d b(\omega,T)}{d T}\biggr) \Delta T. 
\end{equation}
Here, $b(\omega,T)$ is the Bose-Einstein distribution function for phonons with frequency $\omega$ and temperature $T$. 

It remains to determine $T_{ph}(\omega)$, which can be done with the help of phonon Non-equilibrium Green's Functions (NEGF) \cite{Mingo2006}. As input parameter, the dynamical matrix of the CBRAM cell must be constructed. The frozen phonon approach \cite{Togo2015} is the method of choice to generate the dynamical matrix of such large, disordered systems. It relies on the displacement of individual atoms and the evaluation of the first derivative of the forces acting on each atom. A large number of device configurations with a single displaced atom must be constructed and simulated, which is computationally very expensive if done at the \emph{ab initio} level. Therefore, we created for our Ag/a-SiO$_2$ system a force-field parameter set capable of producing lattice dynamics and inter-atomic forces close to DFT accuracy, but at much lower computational cost. Since conventional force-fields face severe difficulties at describing metal/oxide interfaces at sufficiently high accuracy, we trained a machine-learned moment-tensor-potential (MTP) \cite{Shapeev2016} using the Synopsys QuantumATK S-2021.06 software \cite{Smidstrup2020,Schneider2017}. The MTP was trained based on a combination of bulk Ag and SiO$_2$ data and AIMD trajectories of different interfaces, clusters, and filaments, extended by active learning simulations, as described in more details in Appendix \ref{app:forcefield}. Particular emphasis was set on the reproduction of the phonon properties of Ag and SiO$_2$. The dynamical matrix produced by QuantumATK was then passed to OMEN to perform the phonon transport calculations \cite{Rhyner2014}. 

\begin{figure}
\includegraphics[width=8.6cm]{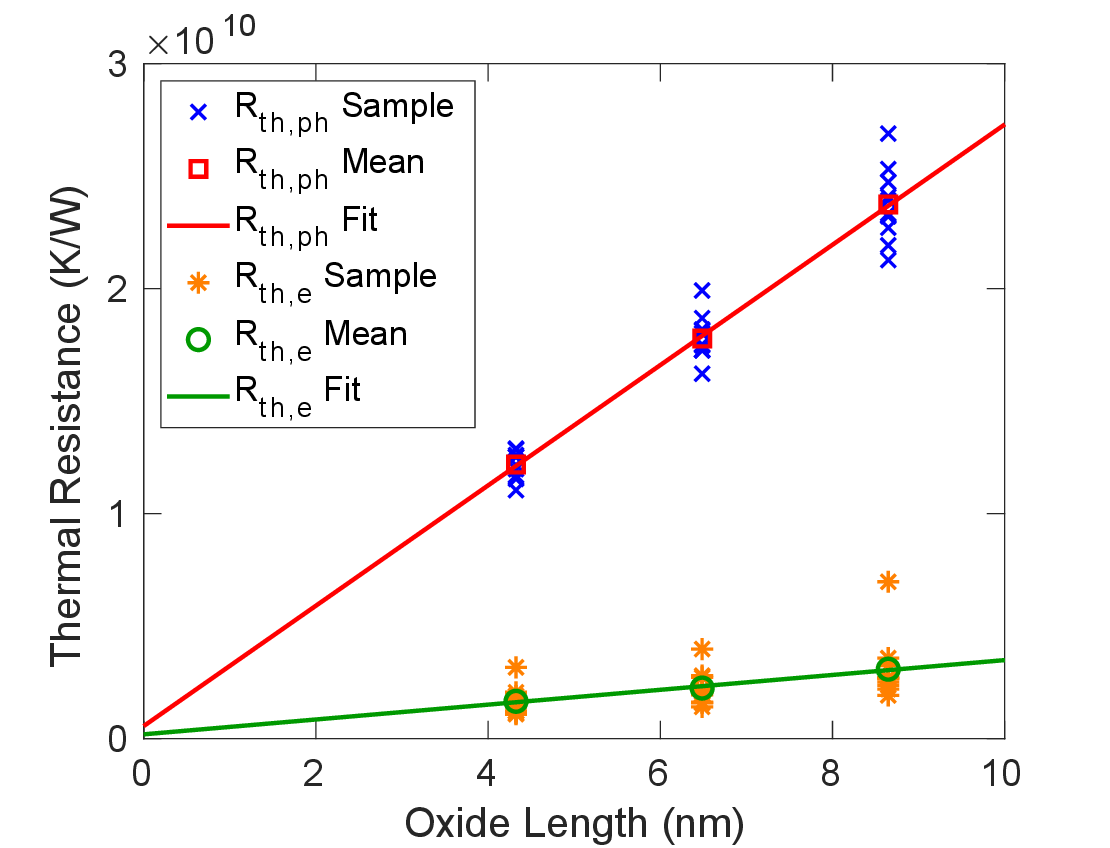}
\caption{\label{fig:thermalcond} Electron and phonon contributions to the thermal resistance as a function of the CBRAM oxide length. Ten Ag/a-SiO$_2$/Ag samples as the one in Fig.~\ref{fig:fig_cond_theory}(a) were evaluated with three different oxide lengths. The stars represent the electronic contribution to the thermal resistance ($R_{th,e}$) of each individual sample, the circles are the mean values at each length, and the solid green line is a linear fit through the three circles. Similarly, the crosses represent the phononic contribution to the thermal resistance ($R_{th,ph}$) of each individual sample, the squares are the mean values at each length, and the solid red line is a linear fit through the three squares. }
\end{figure}

The thermal conductance components $G_{th,e}$ and $G_{th,ph}$ were calculated at the same device lengths as their electrical counterparts. The corresponding thermal resistances $R_{th,e} = 1/G_{th,e}$ and $R_{th,ph} = 1/G_{th,ph}$ are reported in Fig.~\ref{fig:thermalcond}. Here again, a linear increase of these resistances can be observed with increasing oxide thickness. This fact is key to extract the thermal resistivities 
\begin{equation}
    \varrho_{th,e} = \frac{dR_{th,e}}{dL},
\end{equation}
and
\begin{equation}
    \varrho_{th,ph} = \frac{dR_{th,ph}}{dL},
\end{equation}
and the two thermal conductivities 
\begin{equation}
    \kappa_{th,e} = \frac{1}{\varrho_{th,e}A} = 7.48~\text{W/(m K)},
\end{equation}
and
\begin{equation}
    \kappa_{th,ph} = \frac{1}{\varrho_{th,ph}A} = 0.97~\text{W/(m K)}. 
\end{equation}
In the above equations, $A$ is the filament cross section area. Obviously, the electronic contribution is much larger than the phonon one indicating that the Ag filament keeps its metallic character. The total thermal conductivity $\kappa$ is finally given by 
\begin{equation}
    \kappa = \kappa_{th,e} + \kappa_{th,ph}, 
\end{equation}
which is equal to 8.45~W/(m K) for the simulated Ag filament. Note that the bulk thermal conductivity of Ag is 50 times larger than this value ($\kappa_{\text{Ag, bulk}} = 429$~W/(m K) \cite{Ho1972}). For the thermal conductivity of the bulk Ag and Pt electrodes and of the a-SiO$_2$ layer, the experimental values from the literature \cite{Kleiner1996,Ho1972} were retained. A summary of all parameters used as inputs to the FEM model is given in Table~\ref{tab:table_values}, together with their origin. 

\begin{table*}
\caption{\label{tab:table_values}
Summary of the input parameters used in the Ag/a-SiO$_2$/Pt CBRAM FEM model. They are divided into five categories: diffusion, electrical transport, chemical processes, thermal properties, and general device/experimental parameters. The values are either derived from \emph{ab initio}/semi-empirical calculations ($\bigstar$), fixed based on experiments ($\bigcirc$), or fitted ($\lozenge$). The chemical parameters derived from \emph{ab initio} simulations had to be increased by 0.05~eV (max.\ 10\% change) to guarantee the numerical stability of the FEM model ($\blacksquare$). }
\begin{ruledtabular}
\begin{tabular}{lcllcl}
\textbf{Diffusion}&$\bigstar$&$\Delta E_D$&Energy barrier for diffusion&$0.196$&eV\\
&$\bigstar$&$D_0$&Diffusion at $\infty$ temperature&$2.96 \cdot 10^{-5}$&cm$^2$/s\\
\hline
\textbf{Electrical transport}&$\bigstar$&$C$&Fitting factor for Simmons equation&$5.1$&\\
&$\bigstar$&$\Delta E_\text{tu}$&Tunneling barrier height&$1.03$&eV\\
&$\bigcirc$&$m_\text{eff}$&Effective electron mass&$0.63$&$m_0$\footnote{$m_0$: electron's mass}\\
\hline
\textbf{Chemical processes}&$\blacksquare$&$\Delta E_a$&Energy barrier for oxidation and dissolution&$1.85$\footnote{$\Delta E_a$ derived from \emph{ab initio} calculations: 1.80 eV}&eV\\
&$\blacksquare$&$\Delta E_c$&Energy barrier for reduction and deposition&$0.49$\footnote{$\Delta E_c$ derived from \emph{ab initio} calculations: 0.44 eV}&eV\\
\hline
\textbf{Thermal properties}&$\bigstar$&$\sigma_\text{filament}$&Electrical conductivity Ag filament in a-SiO$_2$&$1.18 \cdot 10^6$&S/m\\
&$\bigstar$&$\kappa_\text{filament}$&Thermal conductivity Ag filament in a-SiO$_2$&$8.45$&W/(m K)\\
&$\bigcirc$&$\kappa_\text{Ag}$&Thermal conductivity bulk Ag&$429$&W/(m K)\\
&$\bigcirc$&$\kappa_\text{Pt}$&Thermal conductivity bulk Pt&$71.6$&W/(m K)\\
&$\bigcirc$&$\kappa_\text{SiO$_2$}$&Thermal conductivity bulk SiO$_2$&$1.1$&W/(m K)\\
&$\bigcirc$&$C_\text{p,Ag}$&Specific heat capacity Ag&$235$&J/(K kg)\\
&$\bigcirc$&$C_\text{p,Pt}$&Specific heat capacity Pt&$133$&J/(K kg)\\
&$\bigcirc$&$C_\text{p,SiO$_2$}$&Specific heat capacity SiO$_2$&$730$&J/(K kg)\\
\hline
\textbf{General parameters}&$\lozenge$&$C_h$&Helmholtz layer capacitance&$10^{-5}$&F/m$^2$\\
&$\lozenge$&$c_\text{OH$^-$}$&OH$^-$ concentration&$0.3$&mol/m$^3$\\
&$\bigcirc$&$M_\text{Ag}$&Molar mass of Ag&$107.87$&g/mol\\
&$\bigcirc$&$\rho_\text{Ag}$&Density of Ag&$10.49$&g/cm$^3$\\
&$\bigcirc$&$\rho_\text{Pt}$&Density of Pt&$21.45$&g/cm$^3$\\
&$\bigcirc$&$\rho_\text{SiO$_2$}$&Density of a-SiO$_2$&$2.20$&g/cm$^3$\\
&$\bigcirc$&$T$&Temperature&$293$&K\\
&$\bigcirc$&$\alpha$&Electron transfer coefficient&$0.5$&\\
&$\bigcirc$&$n$&Number of transferring electrons&$1$&\\
&$\bigcirc$&$\varepsilon_\text{SiO$_2$}$&Relative permittivity of SiO$_2$&3.9&\\
\end{tabular}
\end{ruledtabular}
\end{table*}

\section{\label{sec:results}Results}

\subsection{\emph{I-V} Characteristics}\label{sec:result_iv}

To validate the proposed FEM model with \emph{ab initio} inputs, the simulated "current vs.\ voltage" (\emph{I-V}) characteristics of the Ag/a-SiO$_2$/Pt CBRAM cell from Ref.~\cite{Emboras2018} is compared to experimental data. The device structure consists of an Ag and Pt electrode and a 20~nm thick layer of a-SiO$_2$ in between. The cell was fabricated on a silicon-on-insulator (SOI) wafer. The two metal electrodes were deposited by e-beam evaporation, followed by a lift-off process. All layers were patterned using e-beam lithography. The a-SiO$_2$ was grown through plasma-enhanced chemical vapor deposition (PECVD) at 300$^{\circ}$~C. The active switching region was confined by including a 3-D tapered Si wave guide. 

\begin{figure*}
\includegraphics[width=17.6cm]{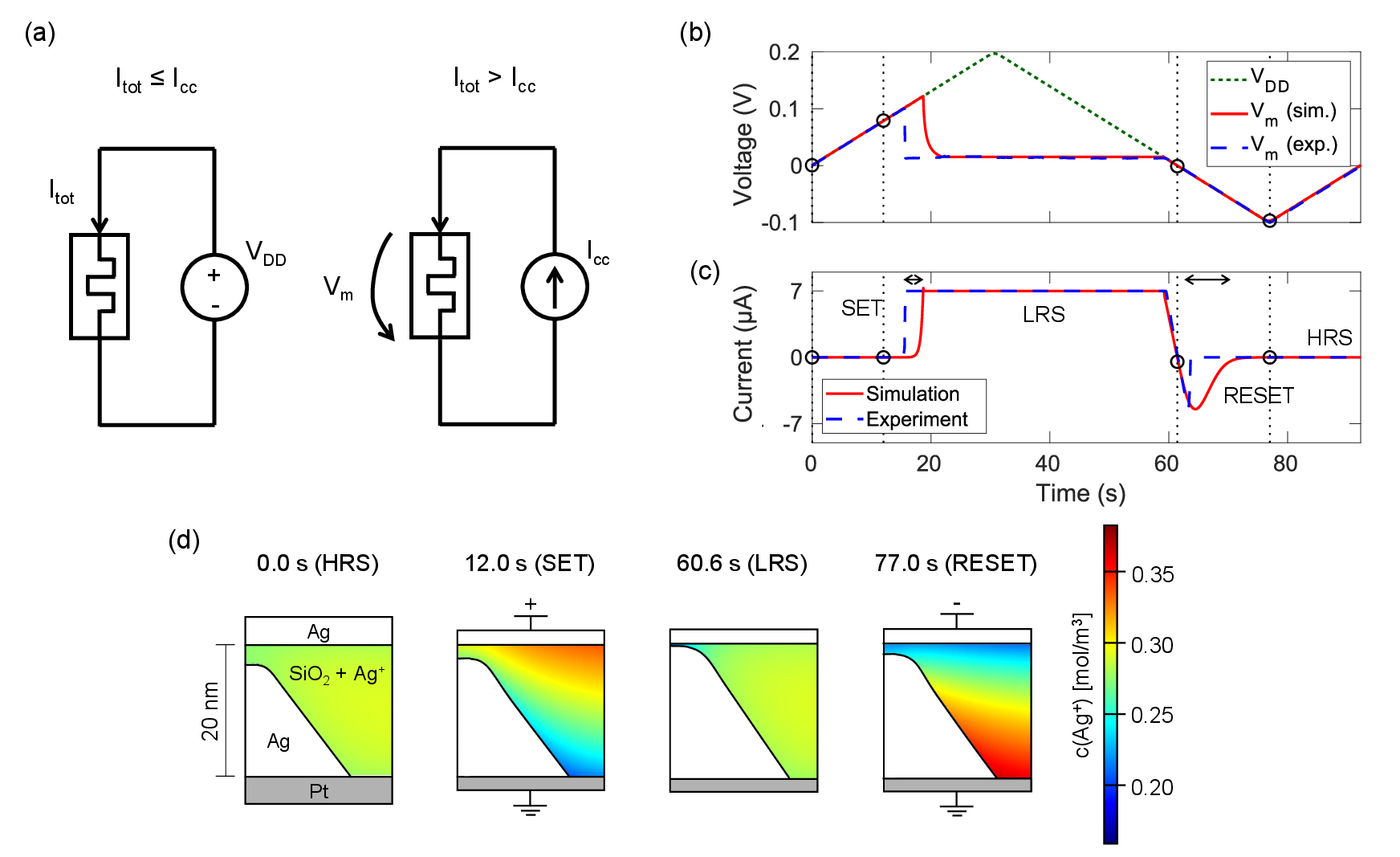}
\caption{\label{fig:fig_comsol} (a) Equivalent electrical circuit measurement setup for the considered Ag/a-SiO$_2$/Pt cell. A voltage $V_\textit{DD}$ is applied to the CBRAM device. It follows the sequence 0 V $\rightarrow$ $+0.2$ V $\rightarrow$ $-0.1$ V $\rightarrow$ 0 V with a sweep rate of 6.5 mV/s. If the total current $I_\text{tot}$ reaches the compliance current $I_{cc}=7$~\textmu A, the circuit representation changes to a current controlled cell. There, the potential drop over the CBRAM is no longer the applied voltage, but the measurement voltage $V_m$. (b) "Voltage vs.\ time" (\emph{V-t}) and (c) "current vs.\ time" (\emph{I-t}) characteristics of the experimental CBRAM (dashed blue line) and the simulated device (solid red line). The largest deviations between the measured and simulated currents are indicated by double arrows. (d) Spatial distribution of the Ag$^+$ ion concentration within the a-SiO$_2$ layer ($c_\text{Ag$^+$}$) at four different times in sub-plots (b) and (c) (marked therein with black circles): at the start (HRS) of the hysteretic cycle (0 s, at $V_\textit{DD}$ = 0 V), during the SET process (12.0 s, at $V_\textit{DD}$ = 0.08 V), in the LRS (60.6 s, at $V_\textit{DD}$ = 0.0 V), and during the RESET process (77.0 s, at $V_\textit{DD}$ = $-0.1$ V). The OH$^-$ background concentration was set to $c_{\text{OH}^-} = 0.3$~mol/m$^3$. Ag$^+$ concentration gradients are established during the SET and RESET processes. }
\end{figure*}

The implemented electrical circuit is illustrated in Fig.~\ref{fig:fig_comsol}(a). As excitation a triangular DC voltage sequence $V_{DD}$ with a voltage amplitude going from $+0.2$~V to $-0.1$~V and a sweep rate of 6.5~mV/s is applied to the CBRAM device. When the total current $I_\text{tot}$ reaches the compliance current $I_{cc} = 7$~\textmu A, the circuit changes to a current controlled cell. There, the potential drop over the CBRAM is no longer $V_{DD}$, but the measurement voltage $V_m$.  

Switching cycles with the same measurement procedure as in the experiment were simulated with the model of Sec.~\ref{sec:approach} and the geometry depicted in Fig.~\ref{fig:fig_model}(a). A CBRAM cell as close as possible to the experimental stack was created, i.e., using a 20~nm thick a-SiO$_2$ layer as solid electrolyte as in the experiment. Figs.~\ref{fig:fig_comsol}(b)-(d) depict the "voltage vs.\ time" (\emph{V-t}) and "current vs.\ time" (\emph{I-t}) characteristics of the full equilibrated SET/RESET cycle for both the modeled and experimental cell, together with the calculated spatial distribution of the Ag$^+$ concentration at four selected times. Starting from a pristine Ag/a-SiO$_2$/Pt CBRAM cell, Ag$^+$ ions are generated at the anode and dissolved in the a-SiO$_2$ layer upon application of a positive voltage to the active electrode. At the same time, OH$^-$ ions are introduced into the oxide up to the predefined average concentration of 0.3~mol/m$^3$ to maintain charge neutrality ($t=0.0$~s in Fig.~\ref{fig:fig_comsol}). The dissolved Ag$^+$ ions get reduced on the filament seed such that the gap between the growing filament and the active electrode continuously decreases according to Eq.~(\ref{eq:vdep})  ($t=12.0$~s in Fig.~\ref{fig:fig_comsol}). Once the gap is short enough to establish an electrical bridge, i.e., when the applied voltage $V_{DD}$ approaches 0.1~V (experiment) and 0.12~V (simulation) in the SET process, the measured current abruptly increases. This step-like behavior is due to the exponential relationship between the tunneling current and the oxide gap, which enables the current to rise by several orders of magnitude until the current compliance is reached. Under current control (when the device is in its LRS ($t=60.6$~s in Fig.~\ref{fig:fig_comsol})) the measured voltage $V_m$ drops to below 0.02~V and the filament geometry does not change anymore. By applying a negative voltage, the reaction rate $r$ changes its direction and filament dissolution sets in. At $V_{DD}=-0.02$~V, the filament has dissolved enough so that the measured current rapidly goes back to very low values although the voltage keeps decreasing. The filament continues to dissolve during the whole time where the voltage $V_{DD}$ is negative ($t=77.0$~s in Fig.~\ref{fig:fig_comsol}). The modeled \emph{V-t} and \emph{I-t} curves agree very well with the experimental measurements with equivalent HRS and LRS values and similar SET/RESET voltages, thus validating the presented model. 

The simulated \emph{I-V} characteristics exhibit a slower switching between the two resistance states than in experiments, as can be seen by the two delays indicated with double arrows in Fig.~\ref{fig:fig_comsol}(c). This discrepancy is a consequence of the assumption of the FEM model that the filament grows according to Eq.~(\ref{eq:vdep}) where the net reaction rate $r$ is the only parameter that determines the growth velocity. Since the filament shape influences the extracted tunneling current through Eq.~(\ref{eq:tunneling}), a continuous change of the gap between the filament tip and the counter electrode results in a steady current variation. This simplification is solely valid at the macroscopic level. In fact, at the atomic level, the a-SiO$_2$ network is heterogeneous and the filament grows discontinuously, which can lead to a sudden increase of the measured current \cite{Onofrio2015,Akola2022,Ducry2017}. Such processes cannot be captured adequately in the chosen continuous model due to its lack of atomic resolution.  

\begin{figure}
\includegraphics[width=8.6cm]{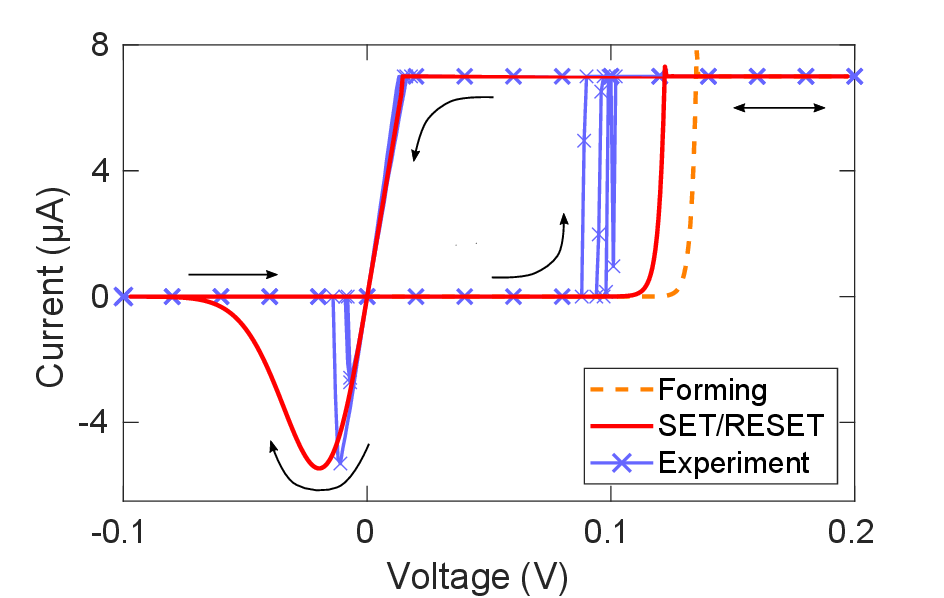}
\caption{\label{fig:fig_iv} "Current vs. voltage" (\emph{I-V}) characteristics of the non-volatile Ag/a-SiO$_2$/Pt CBRAM cell shown in Fig.~\ref{fig:fig_experiment}. Three experimental SET/RESET hysteretic cycles are reported in blue. Individual data points are drawn every 1~mV (experimental sampling resolution) during the SET and RESET process and every 0.02~V otherwise. The compliance current is set to 7~\textmu A. The dashed orange (forming step) and red (SET/RESET) lines represent the simulated \emph{I-V} characteristics with the model of Sec.~\ref{sec:approach}. The arrows indicate the direction of the hysteretic curve. }
\end{figure}

The corresponding \emph{I-V} curve for the considered CBRAM cell is reported in Fig.~\ref{fig:fig_iv}. It features the typical hysteretic behavior expected for non-volatile memory devices. Besides the SET and RESET cycle, the forming step (first SET process) is shown as well. The properties of the forming step primarily depend on the initial filament geometry and dimensions. It follows that the forming voltage is given by the initial gap between the filament tip and the anode. At the same time, the slope of the \emph{I-V} curve around 0~V, when the voltage is ramped down, is indirectly proportional to the resistance of the filament, and, hence, to the shape of the filament tip. Therefore, only the subsequent equilibrated cycles denoted as "SET/RESET" in Fig.~\ref{fig:fig_iv} can be used for comparison with experimental data. In this case the filament has reached a repetitively stable configuration that is independent from the original configuration. 

The energy barriers for the oxidation and dissolution ($\Delta E_a$) and for the reduction and deposition ($\Delta E_c$) in Eq.~(\ref{eq:butlervolmer}) determine the reaction kinetics of the metal/solid electrolyte interfaces. To analyze the impact of these parameters, one can reformulate Eq.~(\ref{eq:butlervolmer}) by substituting $\Delta E_a$ and $\Delta E_c$ by the following quantities
\begin{equation}
    \Delta E_\text{tot} = \frac{\Delta E_a + \Delta E_c}{2}
\end{equation}
and
\begin{equation}
    \Delta E_\text{diff} = \frac{\Delta E_a - \Delta E_c}{2}. 
\end{equation}
Then, Eq.~(\ref{eq:butlervolmer}) can be reformulated as
\begin{eqnarray}
\label{eq:butlerrewrite}
        r = &&\frac{k_B T}{h} \exp{\left(-\frac{\Delta E_\text{tot}}{k_B T} \right)} \nonumber\\
        \Biggl[&&
        \exp{\left(-\frac{\Delta E_\text{diff}}{k_B T}\right)} c_\text{metal} \exp{\left(\frac{\left(1-\alpha\right) n e}{k_B T} \Delta \Phi \right)}\nonumber\\
        -&&\exp{\left(\frac{\Delta E_\text{diff}}{k_B T}\right)} c_\text{ion} \exp{\left(-\frac{\alpha n e}{k_B T} \Delta \Phi \right)}\Biggr].
\end{eqnarray}
Equation~(\ref{eq:butlerrewrite}) illustrates that $E_\text{tot}$ controls the speed during both the SET and RESET process while $E_\text{diff}$ sets the preference for oxidation or reduction. In this model $E_\text{diff}$ only plays a subsidiary role due to the enforced charge neutrality within the SiO$_2$ switching layer. The charge neutrality condition states that the number of dissolved Ag$^+$ ions in the solid electrolyte remains constant after the initialization process. Each Ag$^+$ ion that is reduced to Ag and deposited on the cathode is replaced by another Ag$^+$ ion coming from the anode. Therefore, the kinetics is limited by either the cathodic or the anodic rate. Since a truncated filament must remain available after a full SET/RESET cycle as a seed for the next SET process, $\Delta E_a$ and $\Delta E_c$ cannot be arbitrarily chosen. We realized that the calculated barrier heights $\Delta E_a$ and $\Delta E_c$ in Sec.~\ref{sec:fluxbarrier} lead to a high dissolution velocity during the RESET process resulting in numerical instabilities. Therefore, the extracted $\Delta E_a$ and $\Delta E_c$ values had to be increased by 0.05~eV. This modification increases $\Delta E_\text{tot}$ by also 0.05~eV and hence reduces $r$ (and $v_\text{dep}$ through Eq.~(\ref{eq:vdep})) such that a sufficiently large filament seed remains after the RESET process.  

According to Table~\ref{tab:table_values}, the OH$^-$ concentration $c_\text{OH$^-$}$ and therewith the maximum Ag$^+$ concentration $c_\text{Ag$^+$}$ in the solid electrolyte as well as the Helmholtz layer capacitance $C_h$ are the only free parameters left in the model. A higher Ag$^+$ concentration in the solid electrolyte leads to a smaller potential drop $\Delta \Phi$ over the electrode/solid electrolyte interface. In the absence of large Ag$^+$ concentration gradients in the solid electrolyte, as is the case at high Ag$^+$ concentrations, the electric potential of the solid electrolyte shows similar values at both the anodic and cathodic electrodes so that changing $c_\text{Ag$^+$}$ has only a minimal impact on the switching characteristics of the CBRAM cell. On the contrary, low Ag$^+$ concentrations tend to lead to significant gradients, as can be seen in Fig.~\ref{fig:fig_comsol}(d) for $c_\text{Ag$^+$}=0.3$~mol/m$^3$. There, $c_\text{Ag$^+$}$ is highest at the anode where the Ag$^+$ ions are injected during the SET process, while $c_\text{Ag$^+$}$ becomes smaller in the region close to the bottom of the filament and to the cathode where the Ag$^+$ ions are deposited. Here, the finite diffusion velocity of the Ag$^+$ ions within the a-SiO$_2$ come into play so that the reduced Ag$^+$ ions cannot be replaced, thus establishing a concentration gradient within the electrolyte. This concentration gradient results in slightly different electric potentials in the electrolyte close to the anode and cathode interfaces, which gives rise to slightly less sharp SET processes that take place at lowered voltages. The last free parameter, the Helmholtz capacitance $C_h$, has only a negligible impact on the switching characteristics of the simulated device. When $C_h$ is small, the potential drop occurs mostly over the Helmholtz part of the double layer, leading to slightly sharper HRS-to-LRS transitions during the SET process than with a larger $C_h$.

\subsection{Joule Heating} \label{sec:result_joule}

In the simulation setup outlined in the previous subsection, the ohmic contact between the tip of the Ag filament and the active Ag electrode measures ca.\ 5~nm in diameter in the LRS. Because of the low compliance current of only 7~\textmu A and the thick filament base, which enables heat to be carried away very efficiently, the influence of Joule heating is very limited when the parameters of Sec.~\ref{sec:result_iv} are used in the FEM model. Although the thermal conductivity $\kappa$ and the electrical conductivity $\sigma$ were calculated for ulta-narrow filaments with a diameter of 7~\AA\, the lattice temperature only increases by few Kelvins when Joule heating is turned on. This effect could thus have been neglected here, but this is not always the case. The conditions under which Joule heating becomes relevant and should be absolutely considered in the modeling effort, are discussed in the next paragraph.  

\begin{figure}
\includegraphics[width=7.5cm]{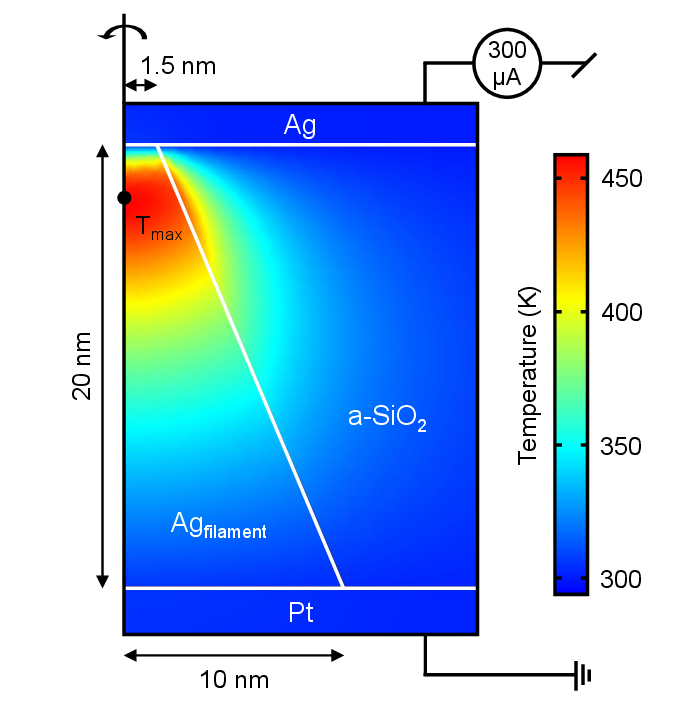}
\caption{\label{fig:joule_setup} 2-D temperature distribution within an Ag/a-SiO$_2$/Pt CBRAM cell similar to the one shown in Fig.~\ref{fig:fig_model}. A bridging, rotationally symmetric, cone-shaped Ag filament of 20~nm height with top and bottom diameters of 3~nm and 20~nm, respectively, is embedded within an a-SiO$_2$ matrix. A current of 300~\textmu A is applied to the Ag top electrode. The temperature is displayed according to the color bar on the right. A maximum temperature of 459~K is measured in the hot spot marked with the black circle labelled with $T_\text{max}$.  }
\end{figure}

To determine geometries and current amplitudes that are subject to Joule heating, Ag/a-SiO$_2$/Pt CBRAM cells in their LRS were constructed with the same dimensions and materials as in Sec.~\ref{sec:result_iv}. Bridging cone-shaped Ag filaments were inserted into the switching layer with a bottom diameter of 20~nm. The diameter of the filament tip in contact with the active electrode was varied between 1~nm and 10~nm. For simplicity, the material parameters from Table~\ref{tab:table_values} were used in all cases, regardless to the filament size. As larger filament geometries are expected to display higher electrical and thermal conductivities, our simulations can be seen as the worst-case scenario. Stationary simulations were then performed for different compliance currents up to 500~\textmu A. Figure~\ref{fig:joule_setup} shows an exemplary temperature distribution for a current of 300~\textmu A and a filament tip diameter of 3~nm. A temperature hot spot of 450~K can be observed in the upper part of the filament, close to the active electrode. This indicates that the thermally well-conducting metallic electrodes efficiently remove the produced heat, contrary to the filament itself and the a-SiO$_2$ matrix, which is known to be a poor thermal conductor. 

\begin{figure}
\includegraphics[width=8.6cm]{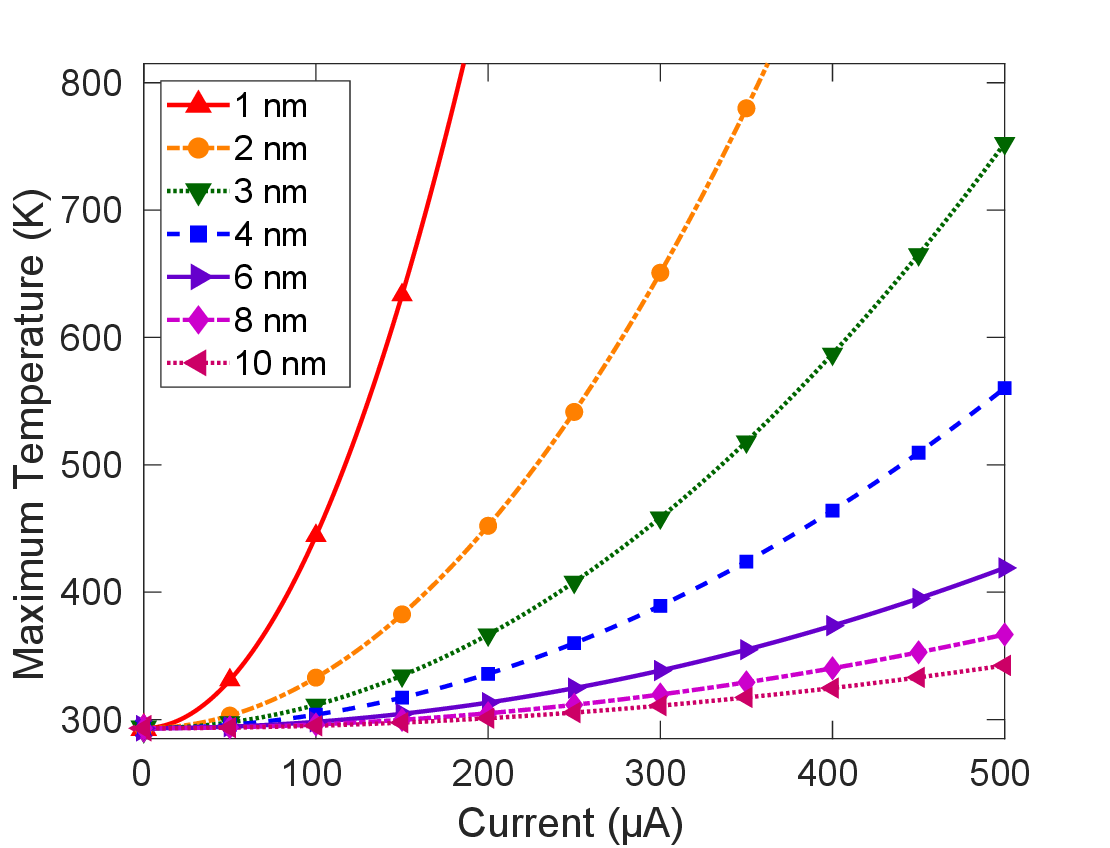}
\caption{\label{fig:joulefilament} Evolution of the maximum temperature measured within Ag filaments embedded in a-SiO$_2$ as a function of their tip diameter and compliance current. The temperature hot spots were determined as indicated in Fig.~\ref{fig:joule_setup}. }
\end{figure}

Assuming that the bridging filament behaves as a perfect resistor and that the input electrical power is fully converted into heat, Joule's law states that
\begin{equation}\label{eq:jouleslaw}
    P_\text{diss} = \frac{I^2}{R}.
\end{equation}
In Eq.~(\ref{eq:jouleslaw}), $P_\text{diss}$ is the dissipated power, $I$ the measured current, and $R$ the filament resistance. Figure~\ref{fig:joulefilament} confirms that this relationship is valid in our devices, further revealing that Joule heating is particularly relevant for cells with narrow filaments (i.e.\ filament diameters below 4-5~nm) and when high currents larger than 100~\textmu A start to flow. 

\section{Conclusions and Outlook} \label{sec:conclusion}

We presented an advanced multiscale framework to simulate the switching  properties of CBRAM cells based on a continuum solver relying on the finite element method. The developed rotationally symmetric 3-D model captures both the growth and dissolution of metallic filaments within an amorphous oxide layer and provides full \emph{I-V} characteristics. The required electronic, ionic, and thermal input parameters are derived from \emph{ab initio} calculations using a broad range of computational materials/device techniques, among which molecular dynamics, nudged elastic band, density functional theory, and quantum transport calculations. To extract these parameters, atomic systems featuring the nanoscale device size and the heterogeneous nature of amorphous oxides were constructed. As key innovation, atomic phonon transport calculations were performed via dynamical matrices obtained from a machine-learned moment-tensor-potential. This allowed to obtain the thermal conductivity of metallic filaments embedded within an oxide. Thanks to the systematic elimination of free parameters, our model overcomes the typically limited predictive capabilities of traditional continuum models. 

In particular, we demonstrated that the modeled \emph{I-V} characteristics of an Ag/a-SiO$_2$/Pt CBRAM cell agree very well with available experimental data. By taking Joule heating into account, we observed that this effect becomes relevant in structures consisting of thin filaments of only few nanometers in diameter and when high currents of at least 100~\textmu A flow through the CBRAM cell. 

The developed tool chain is ready to explore novel, not-yet-fabricated CBRAM-like memristors. Not only could other material stacks be considered, but also structures with different geometries. As an example, the cone-shaped silver filament could be replaced by an hourglass-shaped copper system. Considering the similitude between the principle of operations of CBRAM and valence change memory (VCM) cells, the latter devices could also be investigated with our model. This requires replacing the metallic filament with randomly distributed oxygen vacancies \cite{Clima2014,Larcher2017}. Finally, the created procedure to extract material parameters from representative atomic samples could be used within kinetic Monte Carlo (KMC) models where similar equations with the same input parameters as in FEM models must be solved.


\begin{acknowledgments}

We thank Julian Schneider (Synopsys Inc.) for his help in the parameterization of the moment-tensor-potential force-field. This work was supported by the Werner Siemens Stiftung, by the Swiss National Science Foundation under grant 198612 (ALMOND Sinergia project), and by a grant from the Swiss National Supercomputing Centre (CSCS) under projects s714, s971, and s1119. 

\end{acknowledgments}

\appendix

\section{Parametrization of the Ag/SiO$_2$ Force-Field} \label{app:forcefield}

\begin{figure*}
\includegraphics[width=17.6cm]{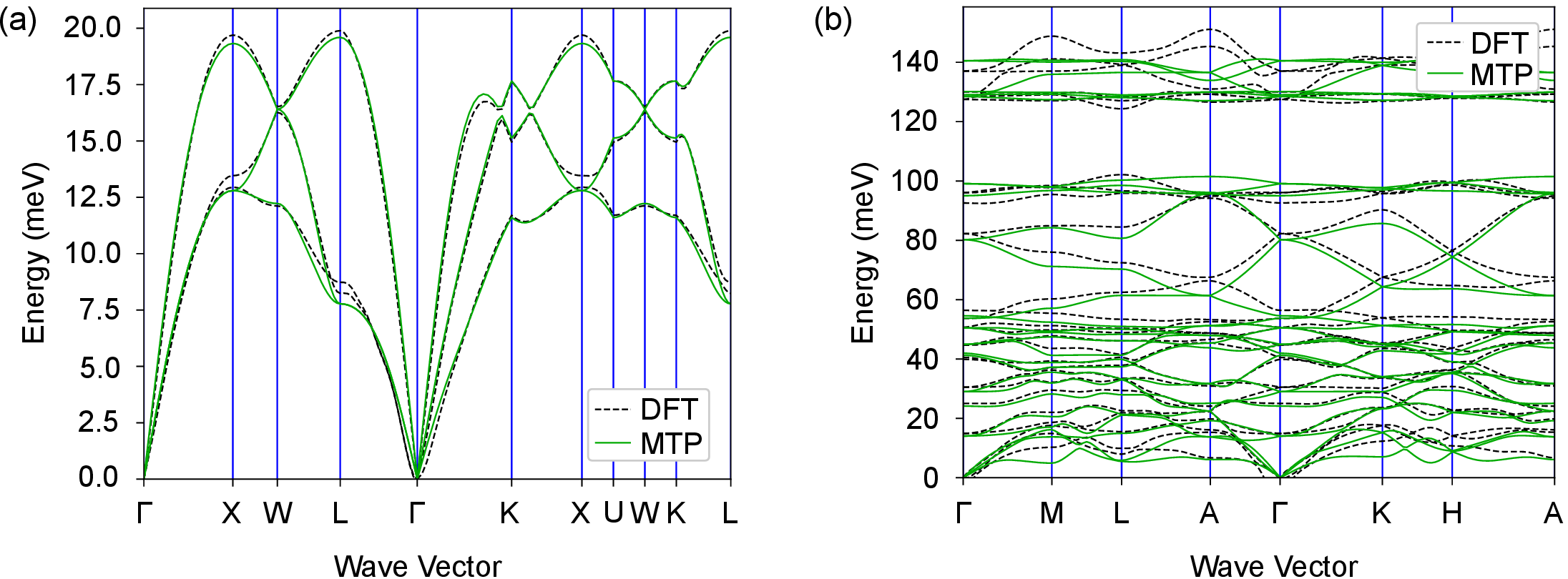}
\caption{\label{fig:fig_bandstructure} Phonon band structures for (a) silver and (b) SiO$_2$ quartz, calculated with DFT (dashed black) and the trained MTP-FF (green). }
\end{figure*}

We trained a machine-learned force field (ML-FF) using the moment-tensor-potential (MTP) approach \cite{Shapeev2016} implemented in the Synopsys QuantumATK software \cite{Smidstrup2020,Schneider2017}.  As reference DFT method we used the linear combination of atomic orbitals (LCAO) approach of QuantumATK. Specifically, we used the pseudo-dojo pseudo-potentials \cite{Vansetten2018} and the corresponding medium basis set with a $k$-point density of 7~\AA. The training data was composed of the following components: bulk crystal silver, bulk crystal silica (quartz and cristobalite), $\langle$111$\rangle$- and $\langle$110$\rangle$-oriented silver/silica (quartz and amorphous) interfaces, as well as silver nanoparticles and nanowires embedded in an amorphous SiO$_2$ matrix. The training data for the bulk materials was generated using random atomic displacements from the equilibrium crystal structures, combined with isotropic and random anisotropic strain on the cell, using the predefined crystal training protocol of QuantumATK. The training data for the interface, nanoparticle, and nanowire systems were generated by first running short AIMD simulations with fast, reduced-accuracy settings at medium temperatures of around 600~K to sample realistic configurations. An initial training data set was then generated by recalculating energies, forces, and stress for equally spaced snapshots from these raw MD trajectories using the final reference DFT calculation method. Finally, for each system the training data was extended by running 100-200~ps long MTP active learning MD simulations at temperatures between 600~K and 1000~K. The active learning simulations start from the previously generated training data and automatically identify new configurations that are not well-represented in the training data based on an extrapolation grade criterion, as explained in Ref.~\cite{Podryabinkin2017}. After calculating DFT energies, forces, and stress, the selected new configurations are added to the training data and the MD simulation is restarted. For the final training all training data sets generated by random displacements, AIMD, and active learning MD are combined in a total training data set. The final MTP was trained using 2000 MTP basis functions sorted according to their level. The training root mean-square error (RMSE) on energies and forces was 11~meV/atom and 0.23~eV/\AA, respectively. The RMSE on an independent test set were of equal magnitude. We validated that the resulting MTP model is stable in MD simulations of various silver/silica interfaces at low and elevated temperatures up to 1000~K. Additionally, we took snapshots from these simulations and recalculated energies and forces with DFT and compared them to the MTP predictions. In all cases we observed RMSE values comparable to the training errors. As further test to validate the applicability of the model with respect to the final target, we compared the phonon band structure of bulk silver and quartz calculated with the reference DFT method and the trained MTP. The results are shown in Fig.~\ref{fig:fig_bandstructure}. For a simple crystal such as silver the phonon spectra are nearly identical. Even for the more complex quartz crystal, which has a more complicated phonon band structure, the agreement between DFT and MTP remains within an acceptable range.

\providecommand{\noopsort}[1]{}\providecommand{\singleletter}[1]{#1}%

\end{document}